\documentclass [aps,prb,showpacs,amsfonts,eqsecnum,twocolumn]{revtex4}

\usepackage{epsfig}
\usepackage{bm}

\newcommand{\smfrac}[2]{\mbox{$\frac{#1}{#2}$}}

\begin{document}

\title{Spin-polarized electron transport in ferromagnet/semiconductor
heterostructures:\ Unification of ballistic and diffusive transport}

\author{R.\ Lipperheide}
\author{U.\ Wille}
\email{wille@hmi.de}

\affiliation{Abteilung Theoretische Physik, Hahn-Meitner-Institut
Berlin,\\ Glienicker Stra\ss{}e 100, D-14109 Berlin, Germany}

\date{\today}

\begin{abstract}

A theory of spin-polarized electron transport in ferromagnet/semiconductor
heterostructures, based on a unified semiclassical description of ballistic and
diffusive transport in semiconductor structures, is developed.  The aim is to
provide a comprehensive framework for systematically studying the interplay of
spin relaxation and transport mechanism in spintronic devices.  A key element
of the unified description of transport inside a (nondegenerate) semiconductor
is the thermoballistic current consisting of electrons which move ballistically
in the electric field arising from internal and external electrostatic
potentials, and which are thermalized at randomly distributed equilibration
points.  The ballistic component in the unified description gives rise to
discontinuities in the chemical potential at the boundaries of the
semiconductor, which are related to the Sharvin interface conductance.  By
allowing spin relaxation to occur during the ballistic motion between the
equilibration points, a thermoballistic spin-polarized current and density are
constructed in terms of a spin transport function.  An integral equation for
this function is derived for arbitrary profile of the electrostatic potential
and for arbitrary values of the momentum and spin relaxation lengths.  Detailed
consideration is given to field-driven spin-polarized transport in a
homogeneous semiconductor.  An approximation is introduced which allows one to
convert the integral equation into a second-order differential equation that
generalizes the standard spin drift-diffusion equation.  The spin polarization
in ferromagnet/semiconductor heterostructures is obtained by invoking
continuity of the current spin polarization and matching the spin-resolved
chemical potentials on the ferromagnet sides of the interfaces.  Allowance is
made for spin-selective interface resistances.  Examples are considered which
illustrate the effects of transport mechanism and electric field.

\end{abstract}


\maketitle

\section{Introduction}

Considerable attention has been devoted in the past few years to the study of
spin-polarized electron transport in hybrid nano\-structures composed of
different types of material, such as nonmagnetic or magnetic semiconductors,
normal metals, ferromagnets, and superconductors.  The motivation behind these
efforts derives from the desire to understand the principles of operation, to
assess the performance, and to explore the field of possible applications, of
solid-state devices relying on the control and manipulation of electron spin
(``spintronic devices'').\cite{wol01,gru02,joh02,zut04} Particular emphasis in
spintronics research is currently placed on the study of spin-polarized
transport in heterostructures formed of a nonmagnetic semiconductor and two
(metallic or semiconducting) ferromagnetic
contacts.\cite{zut04,aws02,sch02,sch05} Structures of this kind are considered
promising candidates for future technological applications.  For the actual
design of spin\-tronic devices, a detailed theoretical understanding of
spin-polarized transport in ferromagnet/semiconductor hetero\-structures is
indispensable.  Up to now, a number of pertinent studies have been undertaken,
which mostly rely on the drift-diffusion model.

Schmidt {\em et al.}\cite{sch00} describe the spin polarization by the same
diffusion equation for the chemical potential as used to treat spin-polarized
transport in ferro\-magnet/\-normal-metal
heterojunctions.\cite{joh87,son87,joh88,val93} For a
metallic-ferromagnet/semiconductor heterojunction with a perfect interface (no
interface resistance or spin scattering), they find that, as a consequence of
the large conductance mismatch, the injected current spin polarization is very
low.  Filip {\em et al.}\cite{fil02} and Rashba\cite{ras00} suggest that
efficient spin injection can be obtained by introducing spin-selective
interface resistances, for example, in the form of tunnel barriers.  This idea
is pursued in a number of detailed theoretical investigations in which the
interface resistances are taken into account either phenomenologically by
introducing discontinuities in the chemical potentials at the
interface,\cite{smi01,fer01,ras02a,yuf02,yuf02a}, or explicitly by treating the
Schottky barrier arising from band bending in the interface depletion
region.\cite{alb02,alb03} Yu and Flatt\'{e}\cite{yuf02,yuf02a} have derived a
drift-diffusion equation for the density spin polarization, which allows, in
particular, the effect of applied electric fields to be studied.  A formalism
taking into account the effect of the electron-electron interaction on
spin-polarized transport in metals and doped (degenerate) semiconductors in the
diffusive regime has been introduced by D'Amico and Vignale,\cite{ami02} and
has subsequently \cite{ami04} been generalized to include the effect of applied
electric fields.  Spin injection under conditions where, in the semiconductor,
ballistic transport prevails over drift-diffusion has been studied by
Kravchenko and Rashba\cite{kra03} within a Boltzmann equation approach; they
find that in the absence of spin-selective interface resistances, the Sharvin
interface conductance\cite{sha65} controls the injection efficiency.  Spin
injection across a Schottky barrier, arising from thermionic emission as well
as tunneling injection, has been treated by Shen at al.\cite{she04} within a
Monte Carlo model.  Phase-coherent transport in the ballistic limit has been
studied, e.g., in Refs.~\onlinecite{tan00,kir01,hum01,hun01,mir01,mat02,zwi03}.

It emerges that the theory of spin-polarized electron transport in
ferromagnet/semi\-conductor heterostructures has reached a level of
considerable sophistication.  Nevertheless, we believe that certain aspects of
the semiconductor part of the transport problem require a more detailed,
unified treatment, such as the interplay of spin relaxation and transport
mechanism all the way from the diffusive to the ballistic regime, and the
effects of the spatial variation of the electrostatic potential profile.  In
the present work, we provide a comprehensive framework for systematically
dealing with these aspects.

The starting point is our unified semiclassical description of (spinless)
ballistic and diffusive electron transport in parallel-plane semiconductor
structures,\cite{lip03} in which the idea of a thermoballistic transport
mechanism was introduced.  The latter relies on the concept of a
``thermoballistic current'' inside the semiconducting sample.  This current
consists of electrons which move ballistically in the electric field arising
from internal (built-in) and external electrostatic potentials, and which are
thermalized at randomly distributed equilibration points (with mean distance
equal to the mean free path, or momentum relaxation length) due to coupling to
the background of impurity atoms and phonons.  The current-voltage
characteristic for nondegenerate systems as well as the zero-bias conductance
for degenerate systems are expressed in terms of a reduced resistance; for
arbitrary momentum relaxation length and arbitrary potential profile, the
latter quantity is determined from a resistance function which is obtained as
the solution of an integral equation.  The thermoballistic chemical potential
and current are derived from this solution as well.  The chemical potential
exhibits discontinuities at the boundaries of the semiconductor, which are
related to the Sharvin interface conductance.

In order to develop, within the unified description, a theory of spin-polarized
electron transport in (nondegenerate) semiconductors, we introduce the
``thermoballistic spin-polarized current'' which generalizes the
thermoballistic current of Ref.~\onlinecite{lip03} so as to allow spin
relaxation to take place during the ballistic motion between the equilibration
points.  The thermoballistic spin-polarized current is constructed in terms of
a ``spin transport function'' that determines the spin polarization inside the
semiconductor for arbitrary potential profile and arbitrary values of the
momentum and spin relaxation lengths.  This function satisfies an integral
equation which follows from the balance equation connecting the thermoballistic
spin-polarized current and density.  The appearance of an integral equation in
the unified description of electron transport (with or without account of the
spin degree of freedom) reflects the nonlocal character of the transport across
the ballistic intervals between the equilibration points.  For electron
transport in a homogeneous semiconductor without space charge, driven by an
external electric field, the integral equation for the spin transport function
can be converted, in an approximation that is adequate for demonstrating the
principal effects of the transport mechanism, into a second-order differential
equation that generalizes the standard spin drift-diffusion equation.  The spin
polarization along a ferromagnet/semiconductor heterostructure is obtained by
invoking continuity of the current spin polarization at the interfaces and
matching the spin-resolved chemical potentials on the ferromagnet sides of the
latter, with allowance for spin-selective interface resistances.

As a prerequisite to developing a theory of spin-polarized electron transport
in semiconductors within the unified transport model of
Ref.~\onlinecite{lip03}, we have to modify and complete the formulation given
in that reference.  This will be done in the next section.  In Sec.\ III, the
spin degree of freedom is introduced into the unified description.  The
integral equation for the spin transport function inside a semiconducting
sample is derived, and the generalized spin drift-diffusion equation is
obtained.  Spin-polarized transport in heterostructures formed of a
nonmagnetic, homogeneous semiconductor and two ferromagnetic contacts is
treated in Sec.\ IV.  We demonstrate the procedure for the calculation of the
current and density spin polarizations across a heterostructure in the
zero-bias limit and of the injected spin polarizations for field-driven
transport.  Various examples are considered which illustrate the effects of
transport mechanism and electric field and exhibit the relation of the unified
description to previous descriptions by other authors.  In Sec.\ V, the
contents of the paper are summarized and our conclusions as well as an outlook
towards applications and extensions of the present work are presented.  In the
Appendix, details of the extended formulation of the unified transport model
outlined in Sec.\ II are worked out.

\section{Unification of ballistic and dif\-fus\-ive transport in
semiconductors}

The unified description of (spinless) ballistic and diffusive electron
transport developed in Ref.~\onlinecite{lip03} has yielded, for the
nondegenerate case, the current-voltage characteristic for a semiconducting
sample enclosed between two plane-parallel contacts.  There, the discontinuity
of the chemical potential has been placed at the interface at one or the other
end of the sample; this gave rise to an ambiguity in the behavior of the
chemical potential {\em inside} the sample.  In order, nevertheless, to obtain
a unique current-voltage characteristic, the reduced resistance determining the
latter was subjected to a heuristic symmetrization procedure (see Sec.\ IV.C of
Ref.~\onlinecite{lip03}).

In the following, we extend in a systematic way the unified description by
treating {\em simultaneously} the effects of the two interfaces on an equal
footing.  In this way, we arrive at unique chemical potentials, currents, and
densities inside as well as at the ends of the semiconducting sample.  This is
prerequisite to the study of spin-polarized electron transport in
ferromagnet/semicinductor heterostructures, which is the principal aim of the
present work.  As in Ref.~\onlinecite{lip03}, we work within the semiclassical
approximation, ignoring all coherence effects.

\subsection{Thermoballistic transport}

\begin{figure}[t]
\vspace*{-1.0cm}
\epsfysize=7.5cm
\epsfbox[195 305 400 700]{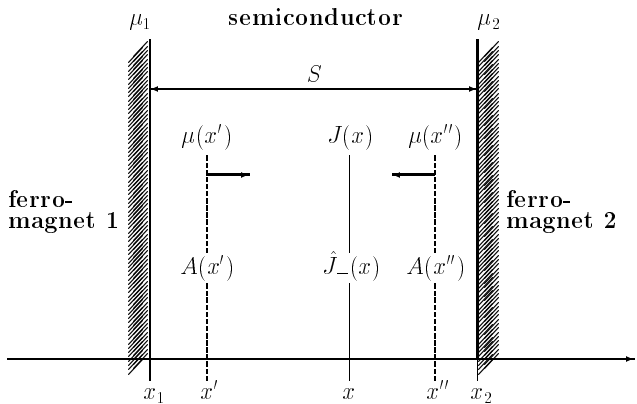}
\vspace*{-0.9cm}
\caption{Schematic diagram showing a semiconducting sample of length $S$
enclosed between two plane-parallel ferromagnetic contacts.  Illustrated are
expression (\ref{eq:11abbc}) for the thermoballistic current $J(x)$ and
expression (\ref{eq:23}) for the off-equilibrium thermoballistic spin-polarized
current $\hat{J}_{-}(x)$. \vspace*{-0.1cm}
}
\label{fig:1}
\end{figure}

We consider a semiconducting sample bordering on a left contact at $x=x_{1}$
and on a right contact at $x=x_{2}$ (see Fig.\ \ref{fig:1}), so that $S =x_{2}
- x_{1}$ is the sample length.  The geometry of the set-up is one-dimensional,
whereas the electron motion is treated in three-dimensional space.
Reformulating the unified description of electron transport in a nondegenerate
semiconductor,\cite{lip03} we introduce the (net) electron current density
(electron current, for short) $J(x',x'')$ across the {\em ballistic interval}
$[x',x'']$ between two equilibration points $x'$ and $x''$,
\begin{eqnarray}
J(x',x'') &=& v_{e} N_{c} \, e^{-\beta E_{c}^{m}(x',x'')} \; \left[ e^{\beta
\mu(x')} - e^{\beta \mu(x'')} \right] \nonumber \\
\label{eq:10xx1}
\end{eqnarray}
($x_{1} \leq x' < x'' \leq x_{2}$), which is the difference of the ballistic
current $J^{l}(x',x'')$ injected at the left end at $x'$,
\begin{equation}
J^{l}(x',x'') = v_{e} N_{c} \, e^{-\beta [E_{c}^{m}(x',x'') - \mu(x')]} \; ,
\label{eq:16xyz1}
\end{equation}
and the analogous ballistic current  $J^{r}(x',x'')$ injected at the right end
at $x''$,
\begin{equation}
J^{r}(x',x'') = v_{e} N_{c} \, e^{-\beta [E_{c}^{m}(x',x'') - \mu(x'')]} \; .
\label{eq:16spd1}
\end{equation}
Here, the function $\mu(x)$ is the chemical potential at the equilibration
point $x$.  [In Ref.~\onlinecite{lip03}, the term quasi-Fermi energy (symbol
$E_{F}$) was used for the chemical potential as defined, e.g., in
Ref.~\onlinecite{ash76}.] Furthermore, $v_{e} = (2\pi m^{*} \beta)^{-1/2}$ is
the emission velocity, $N_{c} = 2(2\pi m^{*}/\beta h^{2})^{3/2}$ is the
effective density of states at the conduction band edge, $m^{*}$ is the
effective mass of the electrons, and $\beta = (k_{B} T)^{-1}$.  The currents
(\ref{eq:16xyz1}) and (\ref{eq:16spd1}) contain only the electrons transmitted
across the sample, i.e., the electrons with sufficient energy to surmount the
potential barrier determined by
\begin{equation}
\hat{E}_{c}^{m}(x',x'') = E_{c}^{m}(x',x'') - E_{c}^{0}
\; ;
\label{eq:20b}
\end{equation}
here, $E_{c}^{m}(x',x'')$ is the maximum value of the potential profile
$E_{c}(x)$ in the interval $[x',x'']$, and $E_{c}^{0}$ is its overall minimum
across the sample.  The profile $E_{c}(x)$ is, in general, a self-consistent
solution of a nonlinear Poisson equation involving the conduction band edge
potential and the external electrostatic potential.

Expressions (\ref{eq:16xyz1}) and (\ref{eq:16spd1}) for the currents injected
at the left and right ends of a ballistic interval follow from Eqs.\ (20) and
(21) of Ref.~\onlinecite{lip03} if classical transmission probabilities are
used.  Tunneling effects can be included by replacing the classical
probabilities with their quantal (WKB) analogues, as done, e.g., in
Ref.~\onlinecite{lip01}.  In the present paper, we do not consider this
possibility.

It is convenient to rewrite Eq.\ (\ref{eq:10xx1}) in the form
\begin{equation}
J(x',x'') = e^{-\beta \hat{E}_{c}^{m}(x',x'')} \; [ {\cal J}(x') - {\cal
J}(x'')] \; ;
\label{eq:10}
\end{equation}
the quantity
\begin{equation}
{\cal J}(x') = v_{e} N_{c} \, e^{- \beta [ E_{c}^{0} - \mu(x')]}
\label{eq:14a}
\end{equation}
is the current injected at the left end $x'$ of the ballistic interval into the
right direction in a flat profile $E_{c}(x) = E_{c}^{0}$, and similarly for the
current ${\cal J}(x'')$ injected at the right end $x''$ into the left
direction.

The ballistic density, at position $x$ in the interval $[x',x'']$, of
transmitted electrons injected at the left end at $x'$ is given by
\begin{eqnarray}
n^{l}(x',x'';x) &=& \frac{N_{c}}{2} \, (2 \beta /\pi m^{\ast})^{1/2} \nonumber
\\ &\ & \times \, \int_{0}^{\infty} dp \, e^{-\beta [p^{2}/2m^{\ast}
+ E_{c}(x) - \mu(x')]} \nonumber \\ &\ & \times \; \Theta \bm{(}p^{2}/2m^{\ast}
+ E_{c}(x) - E_{c}^{m}(x',x'')\bm{)} \nonumber \\ &=& \frac{N_{c}}{2} \,
C(x',x'';x) \, e^{-\beta [E_{c}^{m}(x',x'') - \mu(x')]} \; , \nonumber \\
\label{eq:16yzz1}
\end{eqnarray}
where
\begin{eqnarray}
C(x',x'';x) &=&  e^{\beta [ E_{c}^{m}(x',x'') - E_{c}(x)]} \nonumber \\
&\ & \times \; {\rm erfc}\bm{(}(\beta [ E_{c}^{m}(x',x'') -
E_{c}(x)])^{1/2}\bm{)} \; ; \nonumber \\
\label{eq:dsch1}
\end{eqnarray}
the function ${\rm erfc}(x)$ is the complementary error func\-tion,\cite{abr65}
and $0 < C(x',x'';x) \leq 1$ [the conserved currents (\ref{eq:16xyz1}) and
(\ref{eq:16spd1}) are, of course, independent of $x$]. Analogously, the
ballistic electron density in the transmitted current injected at the right end
at $x''$ is
\begin{eqnarray}
n^{r}(x',x'';x) &=& \frac{N_{c}}{2} \, C(x',x'';x) \, e^{-\beta
[E_{c}^{m}(x',x'') - \mu(x'')]} \; . \nonumber \\
\label{eq:16nico}
\end{eqnarray}
Dividing the current (\ref{eq:16xyz1}) by the density (\ref{eq:16yzz1}) [or
(\ref{eq:16spd1}) by (\ref{eq:16nico})], we obtain for the average velocity
$v(x',x'';x)$ at position $x$ of the electrons injected at either end of, and
transmitted across, the interval $[x',x'']$
\begin{equation}
v(x',x'';x) = \frac{J^{l,r}(x',x'')}{n^{l,r}(x',x'';x)} =
\frac{2 v_{e}}{C(x',x'';x)}  \; ,
\label{eq:16uvw1}
\end{equation}
which depends only on the potential profile.  For constant profile, one has
$C(x',x'';x) = 1$, and the electrons move with the average velocity $2 v_{e}$
from one or the other end to its opposite.  This is the average velocity of the
injected electrons also in the case of a position-dependent profile.  However,
some of these electrons are reflected back, so that the average velocity at
position $x$ of those electrons which have sufficient energy to pass over the
top of the potential profile must be higher than $2 v_{e}$, namely, equal to
the velocity given by Eq.\ (\ref{eq:16uvw1}).

In analogy to Eq.\ (\ref{eq:10xx1}), the sum of the densities $n^{l}(x',x'';x)$
and $n^{r}(x',x'';x)$ is the ballistic density $n(x',x'';x)$ of transmitted
electrons in the interval $[x',x'']$,
\begin{eqnarray}
n(x',x'';x) &=& \frac{N_{c}}{2} \, C(x',x'';x) \, e^{-\beta E_{c}^{m}(x',x'')}
\nonumber \\ &\ & \times \left[ e^{\beta \mu(x')} + e^{\beta \mu(x'')} \right]
\; . \label{eq:10xxx1}
\end{eqnarray}
The density $n(x',x'';x)$, like the current $J(x',x'')$, comprises only
electrons that participate in the transport.

From the ballistic current (\ref{eq:10}), the {\em thermoballistic current}
$J(x)$ at position $x$ inside the semiconductor is constructed by summing up
the weighted contributions of the ballistic intervals $[x',x'']$ for which
$x_{1} \leq x' < x < x'' \leq x_{2}$ [see Eq.\ (23) of
Ref.~\onlinecite{lip03}],
\begin{eqnarray}
J(x) &=& w_{1}(x_{1},x_{2};l) \; [{\cal J}_{1} - {\cal J}_{2}] \nonumber \\ &+&
\int_{x_{1}}^{x}\frac{dx'}{l} \, w_{2}(x',x_{2};l) \, [{\cal J}(x') - {\cal
J}_{2}] \nonumber \\ &+& \int_{x}^{x_{2}} \frac{dx'}{l} \, w_{2}(x_{1},x';l) \,
[{\cal J}_{1} - {\cal J}(x')] \nonumber \\ &+& \int_{x_{1}}^{x} \frac{dx'}{l}
\int_{x}^{x_2} \frac{dx''}{l} \, w_{3}(x',x'';l) \, [{\cal J} (x') - {\cal
J}(x'')] \; , \nonumber \\
\label{eq:11a}
\end{eqnarray}
with
\begin{equation}
w_{n}(x',x'';l)  = p_{n}(|x''-x'|/l) \, e^{-\beta \hat{E}_{c}^{m}(x',x'')}
\label{eq:14}
\end{equation}
$(n=0,1,2,3)$, where $l$ is the momentum relaxation length of the electrons,
which comprises the effect of relaxation due to electron scattering by impurity
atoms and phonons.  The probabilities $p_{n}(x/l)$ of occurrence of the
ballistic intervals depend on the dimensionality of the transport [note that
$w_{n}(x',x'';l)$ is symmetric with respect to an interchange of $x'$ and
$x''$].  In Eq.\ (\ref{eq:11a}), the quantities ${\cal J}_{1,2} = {\cal
J}(x_{1,2})$ are fixed by the chemical potentials $\mu_{1,2} = \mu(x_{1,2})$
{\em on the contact sides} of the interfaces at $x_{1,2}$, i.e., immediately
{\em outside} of the sample,
\begin{equation}
{\cal J}_{1,2} = v_{e} N_{c} \, e^{- \beta (E_{c}^{0} - \mu_{1,2})} \; .
\label{eq:14aa}
\end{equation}
For later convenience, we introduce a symbolic operator $\mathbb{W}(x',x'';l)$
to write expression (\ref{eq:11a}) in the condensed form
\begin{eqnarray}
J(x) &=&  \int_{x_{1}}^{x} \frac{dx'}{l} \int_{x}^{x_2} \frac{dx''}{l} \,
\mathbb{W}(x',x'';l) \,  [{\cal J} (x') - {\cal J}(x'')]  \; , \nonumber \\
\label{eq:11aba}
\end{eqnarray}
which, in view of Eqs.\ (\ref{eq:14a}) and  (\ref{eq:14aa}), may also be
written as
\begin{eqnarray}
J(x) &=&  v_{e} N_{c} \, e^{-\beta E_{c}^{0}} \int_{x_{1}}^{x} \frac{dx'}{l}
\int_{x}^{x_2} \frac{dx''}{l} \, \mathbb{W}(x',x'';l) \nonumber \\ &\ &
\times \left[ e^{\beta \mu(x')} - e^{\beta \mu(x'')} \right] \; .
\label{eq:11abbc}
\end{eqnarray}
Analogously, we introduce the {\em thermoballistic density} inside the
semiconductor, $n(x)$,  as
\begin{eqnarray}
n(x) &=&  \frac{N_{c}}{2} \, e^{-\beta E_{c}^{0}} \int_{x_{1}}^{x}
\frac{dx'}{l} \int_{x}^{x_2} \frac{dx''}{l} \, \mathbb{W}_{C}(x',x'';l;x)
\nonumber \\ &\ & \times \left[ e^{\beta \mu(x')} + e^{\beta \mu(x'')} \right]
\; , \label{eq:11abbd}
\end{eqnarray}
where
\begin{equation}
\mathbb{W}_{C}(x',x'';l;x) = \mathbb{W}(x',x'';l) \,  C(x',x'';x)
\label{eq:11abbs} \; .
\end{equation}
Again, the current $J(x)$ and the density $n(x)$ comprise only electrons that
participate in the transport.

The thermoballistic current (\ref{eq:11a}) by itself is not, in general,
conserved, but together with the {\em background current}\cite{lip03}
$J^{back}(x)$ it adds up to the conserved physical current $J$:
\begin{equation}
J(x) + J^{back}(x) = J ={\rm const.}
\label{eq:3}
\end{equation}
The background current is confined within the sample and, therefore, must
vanish when integrated over the latter, which implies that the thermoballistic
current $J(x)$ averaged over the sample yields the physical current $J$,
\begin{equation}
\frac{1}{x_{2} - x_{1}} \int_{x_{1}}^{x_{2}} dx \, J(x) =  J  \; .
\label{eq:5}
\end{equation}
The non-conservation of the thermoballistic current can be viewed as arising
from a source term $Q(x)$ associated with the gain of thermoballistic electron
density due to the coupling to the background, as expressed by the equation
\begin{equation}
\frac{dJ(x)}{dx} = Q(x) \; .
\label{eq:5a}
\end{equation}
In the background, the source term appears as a sink term describing the loss
of electron density,
\begin{equation}
Q^{back}(x) = - Q(x) \; .
\label{eq:5b}
\end{equation}
Again, since the background electrons are confined to the sample, the integral
of $Q^{back}(x)$ over the sample must vanish and, therefore, also
that of the thermoballistic source term  $Q(x)$,
\begin{equation}
\int_{x_{1}}^{x_{2}} dx \, Q(x) = 0 \; .
\label{eq:5d}
\end{equation}
Owing to Eq.\ (\ref{eq:5a}), this implies
\begin{equation}
J(x_{1}^{+}) = J(x_{2}^{-}) \equiv \kappa \, J \; ,
\label{eq:5e}
\end{equation}
that is, the thermoballistic current entering at one end of the sample, $x =
x_{1}^{+}$, must be the same as the one leaving at the other end, $x =
x_{2}^{-}$. The quantity $\kappa$ has been introduced for later convenience;
it normalizes the thermoballistic current on the sample sides of the
ferromagnet/semiconductor interfaces to the physical current $J$.  We remark
that, in analogy to the thermoballistic current, the thermoballistic density as
well as other thermoballistic quantities introduced later in the development
all have their background complement.

Condition (\ref{eq:5}) has been used in Ref.~\onlinecite{lip03} to obtain the
fundamental integral equation for the determination of the thermoballistic
current.  Condition (\ref{eq:5e}) is new, and provides us with an extension of
the formalism of Ref.~\onlinecite{lip03} which allows us to establish a unique
thermoballistic chemical potential inside the sample, as will be shown in the
following.

\subsection{Thermoballistic chemical potential, current, and density}

Substituting expression (\ref{eq:11a}) in condition (\ref{eq:5}), we obtain
\begin{eqnarray}
\frac{x_{2}-x_{1}}{l} \, J &=& \left[ u_{1}(x_{1},x_{2};l) +
\int_{x_{1}}^{x_{2}} \frac{dx'}{l} \, u_{2}(x_{1},x';l) \right] {\cal
J}_{1} \nonumber \\ &-& \left[ u_{1}(x_{1},x_{2};l) + \int_{x_{1}}^{x_{2}}
\frac{dx'}{l} \, u_{2}(x',x_{2};l) \right] {\cal J}_{2} \nonumber \\
&+& \int_{x_{1}}^{x_{2}} \frac{dx'}{l} \left[ \rule{0mm}{5mm} u_{2}(x',x_{2};l)
- u_{2}(x_{1},x';l) \right. \nonumber \\ &\ & + \left. \int_{x_{1}}^{x_{2}}
\frac{dx''}{l} \, u_{3}(x',x'';l) \right] {\cal J}(x') \; ,
\label{eq:11b}
\end{eqnarray}
where
\begin{equation}
u_{n}(x',x'';l)  = \frac{x''-x'}{l} \, w_{n}(x',x'';l)
\label{eq:14b}
\end{equation}
[note that $u_{n}(x',x'';l)$ is antisymmetric with respect to an interchange of
$x'$ and $x''$]. Equation (\ref{eq:11b}) is a basic condition on the function
${\cal J}(x)$, and hence, via Eq.\ (\ref{eq:14a}), on the chemical potential
$\mu(x)$, whose determination allows all relevant transport quantities to be
obtained.

For given current $J$, only the value of the chemical potential at one of the
interfaces with the contacts can be prescribed.  If, at the interface at
$x_{1}$, we prescribe the value of $\mu_{1}$, i.e., that of ${\cal J}_{1}$
[case (i)], we can find the value of ${\cal J}(x)$ at the other interface at
$x_{2}$ by re-expressing in Eq.\ (\ref{eq:11b}) ${\cal J}_{2}$ as ${\cal
J}(x_{2})$ and replacing $x_{2}$ with the variable $x$, thereby obtaining an
integral equation for the function ${\cal J}(x)$ in the range $x_{1} < x \leq
x_{2}$.  If, now, ${\cal J}(x_{2})$ is required to assume a {\em preassigned}
value for which we re-introduce the symbol ${\cal J}_{2}$, then the current $J$
on the left-hand side of Eq.\ (\ref{eq:11b}) is fixed at some value $J_{1}$.
We denote the associated solution of the integral equation by ${\cal
J}_{1}(x)$.

On the other hand, prescribing the value $\mu_{2}$ for the chemical potential
at the interface at $x_{2}$ [case (ii)], we re-express in Eq.\ (\ref{eq:11b})
${\cal J}_{1}$ as ${\cal J}(x_{1})$.  Then, replacing $x_{1}$ with the variable
$x$, we obtain an integral equation for the function ${\cal J}(x)$ in the range
$x_{1} \leq x < x_{2}$.  With ${\cal J}(x_{1})$ required to assume a {\em
preassigned} value ${\cal J}_{1}$, the current $J$ is now fixed at the value
$J_{2}$.  The associated solution of the integral equation is denoted by ${\cal
J}_{2}(x)$.

To determine the functions ${\cal J}_{1,2}(x)$ explicitly, we proceed in the
following way. In case (i), we define the ``resistance function''\cite{lip03}
\begin{equation}
\chi_{1}(x) = \frac{{\cal J}_{1} - {\cal J}_{1}(x)}{J_{1}} \; ; \;\;
\chi_{1}(x_{1}) = 0  \; ,
\label{eq:6}
\end{equation}
and obtain from Eq.\ (\ref{eq:11b}), following the procedure outlined above,
\begin{eqnarray}
\frac{x-x_{1}}{l} &-& \left[ u_{1}(x_{1},x;l) + \int_{x_{1}}^{x}
\frac{dx'}{l} \, u_{2}(x',x;l) \right] \chi_{1}(x) \nonumber \\ &+&
\int_{x_{1}}^{x} \frac{dx'}{l} \left[ \rule{0mm}{5mm} u_{2}(x',x;l) -
u_{2}(x_{1},x';l) \right. \nonumber \\ &\ & + \left. \int_{x_{1}}^{x}
\frac{dx''}{l} \, u_{3}(x',x'';l) \right] \chi_{1}(x') = 0
\; , \nonumber \\
\label{eq:11bb}
\end{eqnarray}
which is a linear, Volterra-type integral equation for $\chi_{1}(x)$. Letting
$x \rightarrow x_{1}^{+}$ in Eq.\ (\ref{eq:11bb}), we find, using the
properties of $u_{n}(x',x'';l)$,
\begin{eqnarray}
\chi_{1}(x_{1}^{+}) &=& \frac{{\cal J}_{1} - {\cal J}_{1}(x_{1}^{+})}{J_{1}}
= e^{\beta [E_{c}(x_{1}) - E_{c}^{0}]} \neq \chi_{1}(x_{1}) \; . \nonumber \\
\label{eq:8}
\end{eqnarray}
With this discontinuity incorporated in it, the solution $\chi_{1}(x)$ is
unique and continuous for $x_{1} < x \leq x_{2}$.  The solution of Eq.\
(\ref{eq:11bb}) can be obtained in closed form under special conditions; in the
general case, one solves this equation efficiently by discretization and
numerical propagation, using the initial value $\chi_{1}(x_{1}^{+})$ given by
Eq.~(\ref{eq:8}).

In case (ii), we define the resistance function
\begin{equation}
\chi_{2}(x) = \frac{{\cal J}_{2}(x) - {\cal J}_{2}}{J_{2}} \; ; \;\;
\chi_{2}(x_{2}) = 0  \; ,
\label{eq:7}
\end{equation}
which satisfies the integral equation
\begin{eqnarray}
\frac{x_{2}-x}{l} &-& \left[ u_{1}(x,x_{2};l) + \int_{x}^{x_{2}}
\frac{dx'}{l} \, u_{2}(x,x';l) \right] \chi_{2}(x) \nonumber \\ &-&
\int_{x}^{x_{2}} \frac{dx'}{l} \left[ \rule{0mm}{6mm} u_{2}(x',x_{2};l) -
u_{2}(x,x';l) \right. \nonumber \\ &\ & + \left . \int_{x}^{x_{2}}
\frac{dx''}{l} \, u_{3}(x',x'';l) \right] \chi_{2}(x') = 0
\; . \nonumber \\
\label{eq:11bbb}
\end{eqnarray}
The solution $\chi_{2}(x)$ is discontinuous at $x = x_{2}$:\
\begin{eqnarray}
\chi_{2}(x_{2}^{-}) &=& \frac{{\cal J}_{2}(x_{2}^{-}) - {\cal J}_{2})}{J_{2}} =
e^{\beta [E_{c}(x_{2}) - E_{c}^{0}]} \neq \chi_{2}(x_{2}) \; . \nonumber \\
\label{eq:8a}
\end{eqnarray}
It follows from Eqs.\ (\ref{eq:11bb}) and (\ref{eq:11bbb}), using the
properties of $u_{n}(x',x'';l)$, that the functions $\chi_{1}(x)$ and
$\chi_{2}(x)$ are related by
\begin{equation}
\chi_{2}(x) = \chi_{1}^{*}(x_{0} - x) \; ,
\label{eq:11abg}
\end{equation}
where $x_{0} = x_{1} + x_{2}$; the asterisk attached to $\chi_{1}$ indicates
that this function is to be calculated using the reverse of the profile
$E_{c}(x)$, given by $E_{c}^{*}(x) = E_{c}(x_{0}-x)$.  If the profile is
symmetric, $E_{c}^{*}(x) = E_{c}(x)$, the functions $\chi_{1}(x)$ and
$\chi_{2}(x)$ are the reverse of one another, $\chi_{2}(x) =
\chi_{1}(x_{0}-x)$.

The two functions ${\cal J}_{1,2}(x)$ are not, in general, equal and yield
different chemical potentials $\mu_{1,2}(x)$ via Eq.\ (\ref{eq:14a}).  Then, in
view of Eq.\ (\ref{eq:14aa}), Eq.\ (\ref{eq:8}) implies $\mu_{1}(x_{1}^{+})
\neq \mu_{1}$, and the chemical potential $\mu_{1}(x)$ is discontinuous at the
interface at $x = x_{1}$, i.e., its value on the semiconductor side of the
interface is not equal to its value at the interface itself.  Analogously,
$\mu_{2}(x_{2}^{-}) \neq \mu_{2}$.  The ambiguity thus found is a
generalization of the ambiguity of the chemical potential in the ballistic
limit $l/S \rightarrow \infty$,\cite{dat95} where it may either be associated
with the current injected at $x_{2}$, in which case it is discontinuous at $x =
x_{1}$ (Sharvin resistance at the interface at $x = x_{1}$), or with the
current injected at $x = x_{1}$, in which case it is discontinuous at $x =
x_{2}$ (Sharvin resistance at the interface at $x = x_{2}$).  In the Appendix,
details of the construction of a {\em unique} thermoballistic chemical
potential $\mu(x)$, current $J(x)$, and density $n(x)$ in terms of the two
solutions $\chi_{1,2}(x)$ are presented.  Here, we only summarize the results.

A quantity $\chi$ is introduced as
\begin{equation}
\chi = \hat{a}_{1} \chi_{1}(x_{2}) + \hat{a}_{2} \chi_{2}(x_{1}) \; ,
\label{eq:12e}
\end{equation}
where $\hat{a}_{1} + \hat{a}_{2} = 1$. The coefficients $\hat{a}_{1}$ and
$\hat{a}_{2}$ are given by
\begin{equation}
\hat{a}_{1,2} = \frac{a_{1,2}}{a} \; ,
\label{eq:11bbc}
\end{equation}
where
\begin{eqnarray}
a_{1} &=& \int_{x_{1}}^{x_{2}} \frac{dx'}{l} \, \{ w_{2}(x_{1},x';l) \, [
\chi_{2}(x') -  \chi_{2}(x_{1}) ] \nonumber \\ &\ & + \; w_{2}(x',x_{2};l) \,
\chi_{2}(x') \} \; , \label{eq:11bbd}
\end{eqnarray}
\begin{eqnarray}
a_{2} &=&  \int_{x_{1}}^{x_{2}} \frac{dx'}{l} \, \{ w_{2}(x_{1},x';l) \,
\chi_{1}(x') \nonumber \\ &\ & + \; w_{2}(x',x_{2};l) \, [ \chi_{1}(x') -
\chi_{1}(x_{2})] \} \; , \label{eq:11bbe}
\end{eqnarray}
and
\begin{equation}
a = a_{1} + a_{2} \; ;
\label{eq:11bfn}
\end{equation}
for a symmetric potential profile $E_{c}(x)$, we have $\hat{a}_{1} =
\hat{a}_{2} = 1/2$.

The current ${\cal J}(x)$ is expressed as
\begin{equation}
{\cal J}(x) = \frac{1}{2} ({\cal J}_{1} + {\cal J}_{2}) - J \,
\chi_{-}(x) \; ,
\label{eq:12aci}
\end{equation}
where
\begin{eqnarray}
\chi_{-}(x) &=&  \hat{a}_{1} \! \left[ \chi_{1}(x) - \frac{1}{2}
\chi_{1}(x_{2}) \right] - \hat{a}_{2} \! \left[ \chi_{2}(x) - \frac{1}{2}
\chi_{2}(x_{1}) \right]
\nonumber \\
\label{eq:12s}
\end{eqnarray}
$(x_{1} \leq x \leq x_{2})$. The currents ${\cal J}_{1}$ and ${\cal J}_{2}$
satisfy the relation
\begin{equation}
{\cal J}_{1} - {\cal J}_{2} = J \chi \; .
\label{eq:12adac}
\end{equation}
With the use of Eq.\ (\ref{eq:14aa}), the current-voltage characteristic is
then obtained in the form
\begin{equation}
J  = v_{e} N_{c} \, e^{-\beta E_{p}} \, \frac{1}{\tilde{\chi}} \, \left( 1 -
e^{-\beta e V} \right) \; ,
\label{eq:9xy}
\end{equation}
where
\begin{equation}
V = \frac{\mu_{1} - \mu_{2}}{e}
\label{eq:9a}
\end{equation}
is the voltage bias between the contacts, and $E_{p} = E_{c}^{m}(x_{1},x_{2})
- \mu_{1}$; the quantity $\tilde{\chi}$, given by
\begin{equation}
\tilde{\chi} = e^{- \beta \hat{E}_{c}^{m}(x_{1},x_{2})} \, \chi \; ,
\label{eq:12ei}
\end{equation}
is the ``reduced resistance''.\cite{lip03} It replaces, in the present extended
unified description, expression (58) of Ref.~\onlinecite{lip03}, which was
obtained, in a heuristic way, by taking the mean value of the reduced
resistances corresponding to case (i) and (ii), respectively.  To determine
$\tilde{\chi}$ in the diffusive and ballistic regimes, we evaluate the
functions $\chi_{1,2}(x)$ by following the development given in the Appendix of
Ref.~\onlinecite{lip03}.  In the diffusive regime $l/S \ll 1$, we find
$\chi_{1}(x_{2}) = \chi_{2}(x_{1}) = \chi$, which leads to
\begin{equation}
\tilde{\chi} = \frac{1}{2 p_{0}(0)} \int_{x_{1}}^{x_{2}} \frac{dx}{l} \, e^{-
\beta [E_{c}^{m}(x_{1},x_{2}) - E_{c}(x)]} \; ,
\label{eq:12diff}
\end{equation}
where the values of $p_{0}(0)$ for one-, two-, and three-dimensional transport
are given in Sec.\ II of Ref.~\onlinecite{lip03}.  In the ballistic limit $l/S
\rightarrow \infty$, we have $\tilde{\chi} = 1$.

According to Eqs.\ (\ref{eq:14a}), (\ref{eq:14aa}), (\ref{eq:12aci}), and
(\ref{eq:12adac}), the {\em thermoballistic chemical potential} $\mu(x)$ is
given by
\begin{eqnarray}
e^{\beta \mu(x)} &=& \left[ \frac{1}{2} - \frac{\chi_{-}(x)}{\chi}
\right] e^{\beta \mu_{1}}  + \left[ \frac{1}{2} + \frac{\chi_{-}(x)}{\chi}
\right] e^{\beta \mu_{2}} \nonumber \\ &=& \eta_{+} - 2 \,
\frac{\chi_{-}(x)}{\chi} \, \eta_{-}
\label{eq:12ijn}
\end{eqnarray}
$(x_{1} \leq x \leq x_{2})$, where
\begin{equation}
\eta_{\pm} = \frac{1}{2} \left( e^{\beta \mu_{1}} \pm e^{\beta \mu_{2}}
\right)\; ,
\label{eq:eta2}
\end{equation}
and the {\em thermoballistic equilibrium electron density} $\bar{n}(x)$ by
\begin{equation}
\bar{n}(x) = N_{c} \, e^{-\beta [E_{c}(x) - \mu(x)]} \; .
\label{eq:12stu}
\end{equation}
The thermoballistic chemical potential $\mu(x)$ [and hence the thermoballistic
equilibrium density $\bar{n}(x)$] are discontinuous at the interfaces at
$x_{1,2}$,
\begin{equation}
e^{\beta [\mu(x_{1}^{+}) - \mu_{1}]} - 1 = - \hat{a}_{1} \frac{\beta e^{2}J}{
{\cal G}_{1}} \; ,
\label{eq:12stv}
\end{equation}
\begin{equation}
e^{\beta [\mu(x_{2}^{-}) - \mu_{2}]} -1 = \hat{a}_{2}
\frac{\beta e^{2} J}{{\cal G}_{2}} \; ,
\label{eq:12stw}
\end{equation}
as can be shown with the help of Eqs.\ (\ref{eq:8}) and (\ref{eq:8a}),
respectively, and Eqs.\  (\ref{eq:12s}) and (\ref{eq:9xy}). Here,
\begin{equation}
{\cal G}_{1,2} = \beta e^{2} v_{e} \bar{n}_{1,2}
\label{eq:90}
\end{equation}
are the respective Sharvin interface conductances,\cite{sha65,lip03} with
$\bar{n}_{1,2} = \bar{n}(x_{1,2})$.  The discontinuities of the functions
$\exp[\beta \mu(x)]$ and $\bar{n}(x)$ are proportional to $J$. In the diffusive
regime $l/S \ll 1$, where, according to Eqs.\ (\ref{eq:9xy}) and
(\ref{eq:12diff}), $J \propto l/S$, the discontinuities approach zero together
with $l/S$.

The thermoballistic current $J(x)$ is obtained in terms of $\chi_{-}(x)$
and $\chi$ by substituting expression (\ref{eq:12aci}) in combination with Eq.\
(\ref{eq:12adac}) in Eq.\ (\ref{eq:11aba}) [or, more explicitly, in Eq.\
(\ref{eq:11a})],
\begin{eqnarray}
&\ & \hspace*{-0.8cm} J(x) = J \left\{ \rule{0mm}{5mm} w_{1}(x_{1},x_{2};l) \,
\chi \right. \nonumber \\ &+& \int_{x_{1}}^{x} \frac{dx'}{l} \,
w_{2}(x',x_{2};l) \, \left[ \frac{\chi}{2} - \chi_{-}(x') \right]
\nonumber \\ &+& \int_{x}^{x_{2}} \frac{dx'}{l} \, w_{2}(x_{1},x';l) \,  \left[
\frac{\chi}{2} + \chi_{-}(x') \right] \nonumber \\ &+& \left.
\int_{x_{1}}^{x} \frac{dx'}{l} \int_{x}^{x_2} \frac{dx''}{l} \, w_{3}(x',x'';l)
\, [ \chi_{-}(x'') - \chi_{-}(x') ] \right\} \; . \nonumber \\
\label{eq:11aha}
\end{eqnarray}
The thermoballistic density $n(x)$ is found in a similar fashion from
Eq.\ (\ref{eq:11abbd}),
\begin{eqnarray}
&\ & \hspace*{-0.7cm} n(x) = \frac{J}{2 v_{e}} \left\{ \rule{0mm}{5mm} \chi \,
\coth(\beta e V/2) \, \mathfrak{W}(x_{1}, x_{2};l;x) \right.  \nonumber \\ &-&
\int_{x_{1}}^{x} \frac{dx'}{l} \, \mathfrak{w}_{2}(x',x_{2};l;x) \left[
\frac{\chi}{2} + \chi_{-}(x') \right] \nonumber \\ &+& \int_{x}^{x_{2}}
\frac{dx'}{l} \,\mathfrak{w}_{2}(x_{1},x';l;x) \left[ \frac{\chi}{2} -
\chi_{-}(x') \right] \nonumber \\ &-& \left.  \! \int_{x_{1}}^{x} \!
\frac{dx'}{l} \int_{x}^{x_2} \! \frac{dx''}{l} \, \mathfrak{w}_{3}(x',x'';l;x)
\, [ \chi_{-}(x'') + \chi_{-}(x') ] \right\}  , \nonumber \\
\label{eq:11ahab}
\end{eqnarray}
where
\begin{eqnarray}
&\ & \hspace*{-2.2cm} \mathfrak{W}(x_{1}, x_{2};l;x) =
\mathfrak{w}_{1}(x_{1},x_{2};l;x)
\nonumber \\ &+& \int_{x_{1}}^{x} \frac{dx'}{l} \,
\mathfrak{w}_{2}(x',x_{2};l;x) \nonumber \\
&+& \int_{x}^{x_{2}} \frac{dx'}{l} \, \mathfrak{w}_{2}(x_{1},x';l;x) \nonumber
\\  &+& \int_{x_{1}}^{x} \frac{dx'}{l} \int_{x}^{x_2} \frac{dx''}{l} \,
\mathfrak{w}_{3}(x',x'';l;x)
\label{eq:11kaki}
\end{eqnarray}
and
\begin{eqnarray}
\mathfrak{w}_{n}(x',x'';l;x) =w_{n}(x',x'';l) \, C(x',x'';x)
\; .
\label{eq:11kaku}
\end{eqnarray}
In the zero-bias limit $V \rightarrow 0$, expression (\ref{eq:11ahab})
reduces to the form $n(x) = \bar{n}_{1} \, \mathfrak{W}(x_{1}, x_{2};l;x)$,
from which the physical meaning of the function $\mathfrak{W}(x_{1},
x_{2};l;x)$ becomes apparent.

In the diffusive regime $l/S \ll 1$, we have
\begin{equation}
e^{\beta \mu(x)} = \eta_{+} -  \frac{ I_{c}(x_{1},x) -
I_{c}(x,x_{2})}{I_{c}(x_{1},x_{2})} \, \eta_{-} \;,
\label{eq:12wasg}
\end{equation}
where
\begin{equation}
I_{c}(x',x'') =  \int_{x'}^{x''} dz \, e^{\beta E_{c}(z)}  \; ,
\label{eq:12wiii}
\end{equation}
and
\begin{eqnarray}
J(x) = J &=& - \frac{\nu}{e} \, \bar{n}(x) \, \frac{d
\mu(x)}{dx} \nonumber \\ &=& - \frac{\nu}{e} \left[ \bar{n}(x) \,
\frac{dE_{c}(x)}{dx} + \frac{1}{\beta} \, \frac{d \bar{n}(x)}{dx} \right] \; ,
\nonumber \\
\label{eq:4wasg}
\end{eqnarray}
where $\nu = 2 e v_{e} \beta \langle l\rangle$ is the electron
mobility, and $\langle l \rangle = p_{0}(0) \, l$ is the effective momentum
relaxation length.\cite{lip03} Integrating Eq.\ (\ref{eq:4wasg}), we retrieve
the current-voltage characteristic (\ref{eq:9xy}) with $\tilde{\chi}$ given by
Eq.\ (\ref{eq:12diff}).  The thermoballistic density $n(x)$ becomes equal to
the equilibrium density $\bar{n}(x)$ given by Eq.\ (\ref{eq:12stu}).

In the ballistic limit $l/S \rightarrow \infty$, we have
\begin{eqnarray}
&\ & \hspace*{-0.65cm} e^{\beta \mu (x)} = \eta_{+} - \left\{ 2 \,e^{-\beta
E_{c}^{m}(x_{1},x_{2})} \right. \nonumber \\ &\ & \times \left. \left[
\hat{a}_{1} \, e^{\beta E_{c}^{m}(x_{1},x)} - \hat{a}_{2} \, e^{\beta
E_{c}^{m}(x,x_{2})} \right] - (\hat{a}_{1}  - \hat{a}_{2}) \right\} \eta_{-}
\; , \nonumber \\
\label{eq:eta1}
\end{eqnarray}
where $\hat{a}_{1}$ and $\hat{a}_{2}$ are to be calculated from Eqs.\
(\ref{eq:11bbc})--(\ref{eq:11bfn}) with
\begin{equation}
\chi_{1}(x) = e^{\beta \hat{E}_{c}^{m}(x_{1},x)}
\label{eq:trala}
\end{equation}
and
\begin{equation}
\chi_{2}(x) = e^{\beta \hat{E}_{c}^{m}(x,x_{2})} \; .
\label{eq:trali}
\end{equation}
Further,
\begin{equation}
J(x) = J = 2 v_{e} N_{c} \, e^{-\beta E_{c}^{m}(x_{1},x_{2})} \, \eta_{-}
\label{eq:1gysi}
\end{equation}
and
\begin{equation}
n(x) = N_{c} \, C(x_{1},x_{2};x) \, e^{-\beta E_{c}^{m}(x_{1},x_{2})} \,
\eta_{+} \; .
\label{eq:1oskar}
\end{equation}
As our model demands, expressions (\ref{eq:1gysi}) and (\ref{eq:1oskar}) are
identical to the original ballistic expressions (\ref{eq:10}) and
(\ref{eq:10xxx1}) with $x' =x_{1}$ and $x'' = x_{2}$.

\subsection{Field-driven transport in a homogeneous semiconductor}

To illustrate the formalism developed so far, we now consider electron
transport in a homogeneous semiconductor without space charge.  The electrons
are assumed to be driven by a (constant) electric field ${\cal E}$ directed
antiparallel to the $x$-axis, i.e., they are moving in a linearly falling
potential of the form
\begin{equation}
E_{c}(x) = E_{c}(x_{1}) - e |{\cal E}| (x-x_{1})  \; ,
\label{eq:38aaaa}
\end{equation}
in which case
\begin{eqnarray}
C(x',x'';x) &\equiv& C\bm{(}\epsilon (x-x') \bm{)} \nonumber \\  &=&
e^{\epsilon (x - x')} \, {\rm erfc}\bm{(}[\epsilon (x - x')]^{1/2}\bm{)} \; ,
\nonumber \\
\label{eq:29csu}
\end{eqnarray}
where $\epsilon = \beta e |{\cal E}|$. The latter quantity is related to the
voltage bias $V$ via $\epsilon S = \beta e V$, so that
\begin{equation}
\bar{n}_{1} =\bar{n}_{2} = \bar{n} \; ,
\label{eq:38bbbb}
\end{equation}
where use has been made of Eqs.\ (\ref{eq:9a}) and (\ref{eq:12stu}).

\begin{figure}[t]
\vspace*{0.0cm}
\epsfysize=8.0cm
\epsfbox[50 143 445 553]{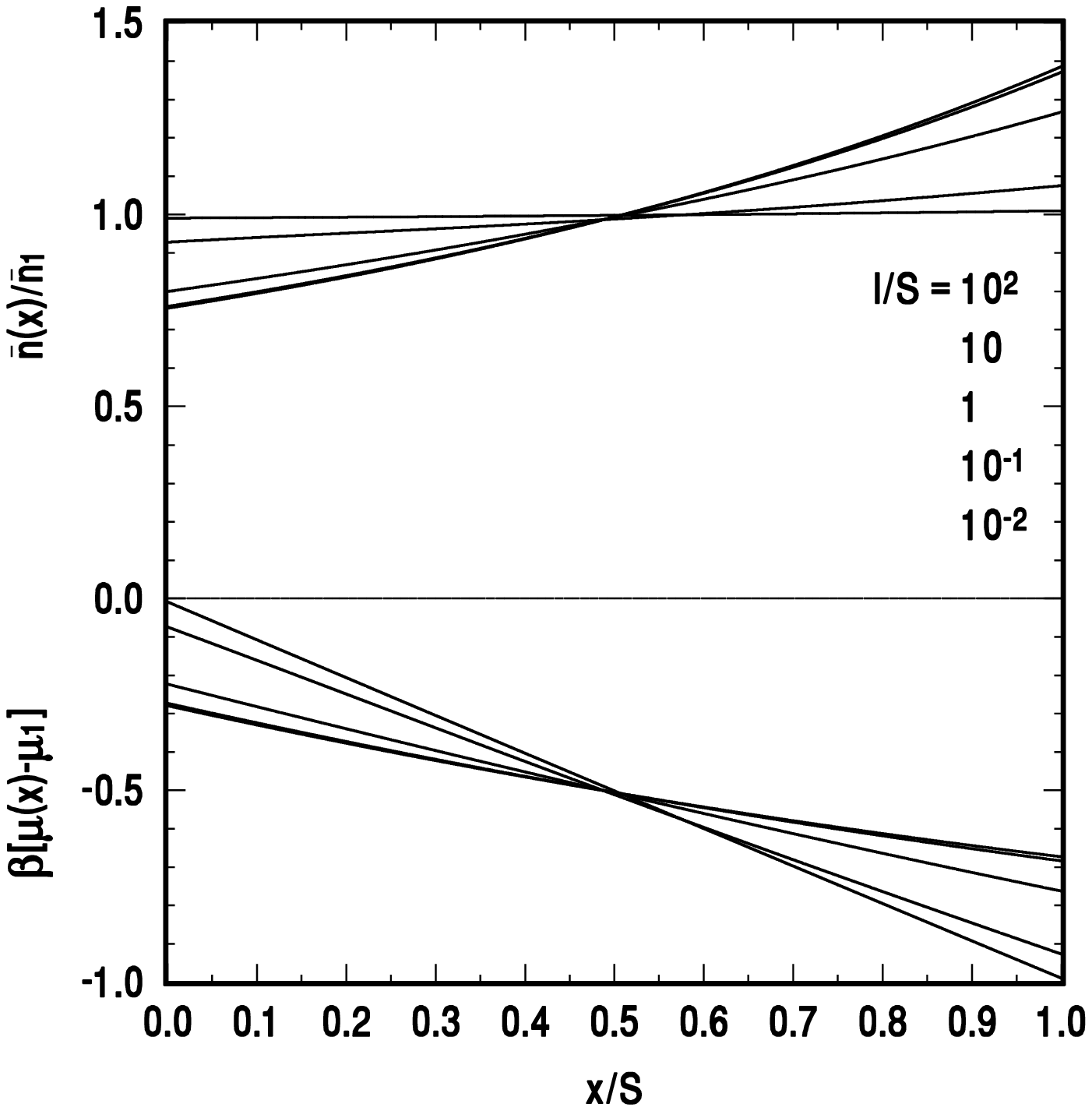}
\vspace*{0.0cm}
\caption{The functions $\beta [ \mu(x) - \mu_{1}]$ and
$\bar{n}(x)/\bar{n}_{1}$, plotted versus $x/S$ for $\epsilon S =1$ and the
indicated values of $l/S$.  The thermoballistic chemical potential $\mu(x)$ is
given by Eq.\ (\ref{eq:12ijn}), and the thermoballistic equilibrium density
$\bar{n}(x)$ by Eq.\ (\ref{eq:12stu}).  The vertical order of the listed $l/S$
values corresponds to the order of the curves within the lower and upper set
of curves, respectively, at $x/S = 1$. \vspace*{-0.3cm}
}
\label{fig:2}
\end{figure}

\begin{figure}[t]
\vspace*{0.0cm}
\epsfysize=8.0cm
\epsfbox[50 143 445 553]{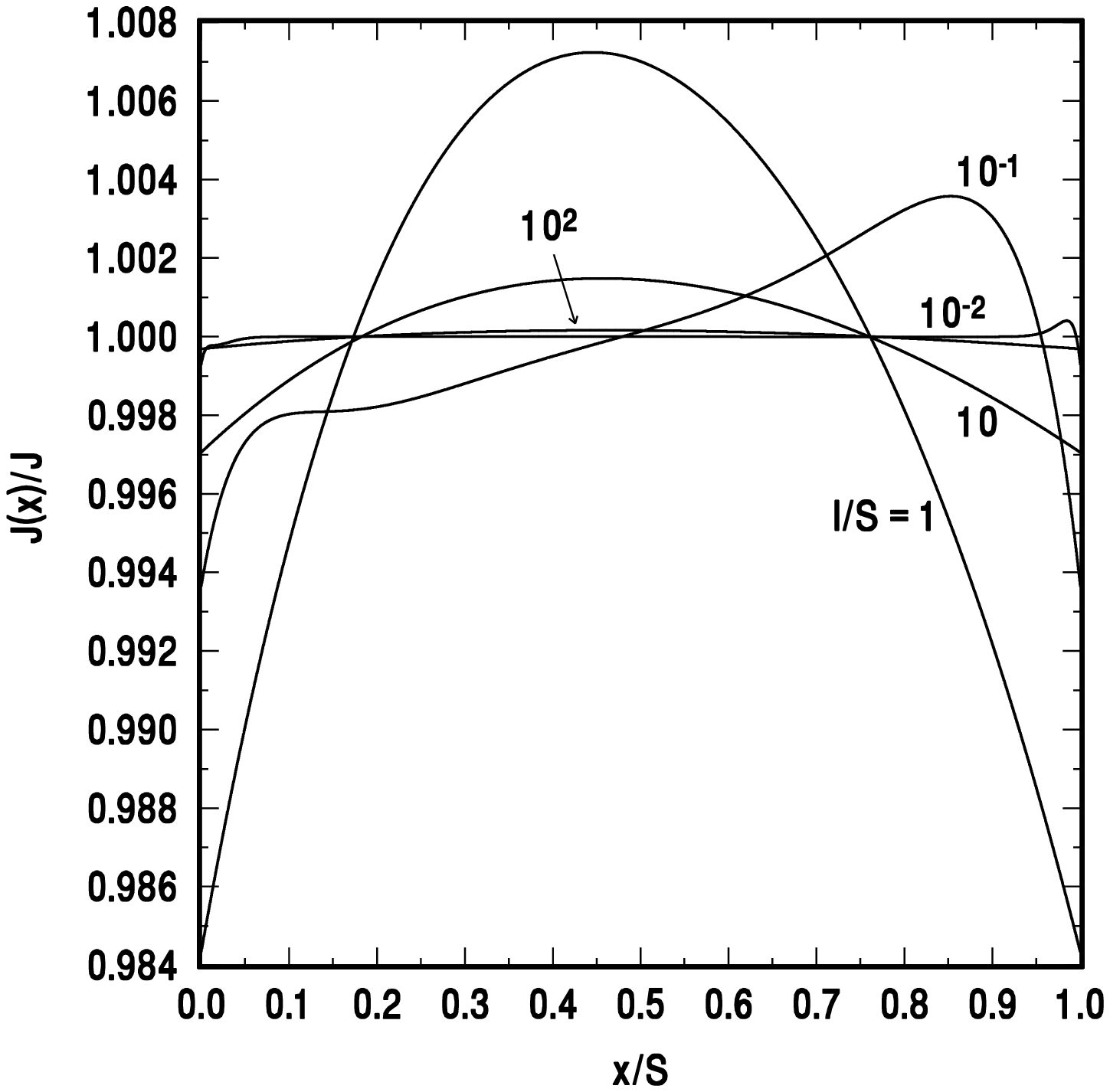}
\vspace*{0.0cm}
\caption{The ratio $J(x)/J$, plotted versus $x/S$ for $\epsilon S =1$ and
the indicated values of $l/S$. The thermoballistic current $J(x)$ is given by
Eq.\ (\ref{eq:11aha}). \vspace*{-0.3cm}
}
\label{fig:3}
\end{figure}

\begin{figure}[t]
\vspace*{0.0cm}
\epsfysize=8.0cm
\epsfbox[50 143 445 553]{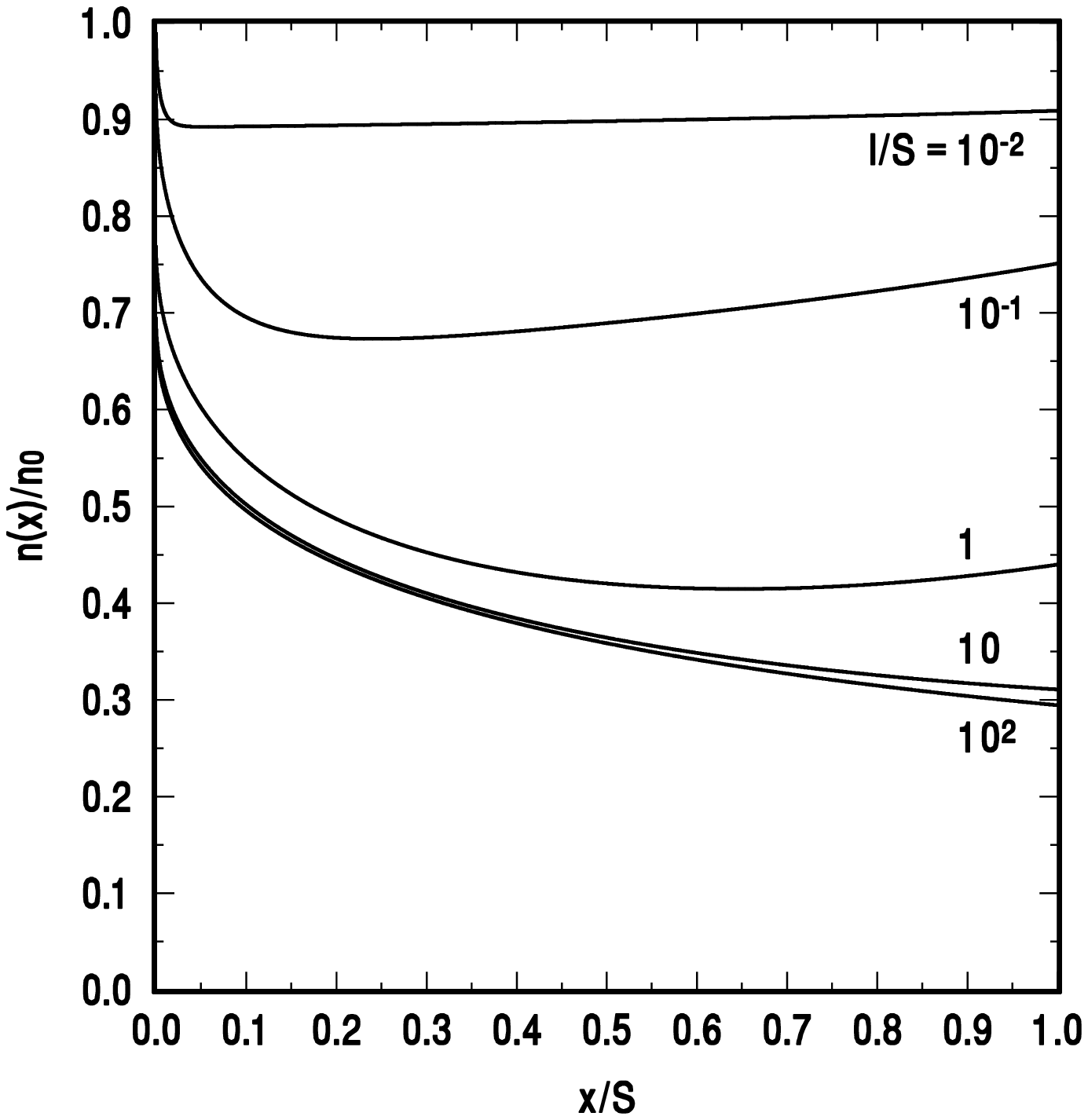}
\vspace*{0.0cm}
\caption{The ratio $n(x)/n_{0}$, plotted versus $x/S$ for $\epsilon S
=1$ and the indicated values of $l/S$. The thermoballistic density $n(x)$ is
given by Eq.\ (\ref{eq:11ahab}), and $n_{0}$ is the constant value that $n(x)$
assumes in the diffusive limit $l/S \rightarrow 0$. \vspace*{-0.3cm}
}
\label{fig:4}
\end{figure}

In order to calculate the thermoballistic chemical potential $\mu(x)$ from Eq.\
(\ref{eq:12ijn}) [or, equivalently, the thermoballistic equilibrium electron
density $\bar{n}(x)$ from Eq.~(\ref{eq:12stu})], the thermoballistic current
$J(x)$ from Eq.~(\ref{eq:11aha}), and the thermoballistic density $n(x)$ from
Eq.~(\ref{eq:11ahab}), we first have to solve the integral equations
(\ref{eq:11bb}) and (\ref{eq:11bbb}) numerically for the functions
$\chi_{1}(x)$ and $\chi_{2}(x)$, respectively.  For convenience, we use in the
integral equations the probabilities $p_{n}(x/l)$ in their one-dimensional
form, $p_{n}(x/l) = e^{-x/l}$ [see Eq.~(10) of Ref.~\onlinecite{lip03}], so
that from Eq.~(\ref{eq:14}) and (\ref{eq:14b})
\begin{eqnarray}
u_{n}(x',x'';l)  &=& \frac{x''-x'}{l} \, e^{-|x''-x'|/l} \, e^{-\epsilon [x_{2}
- \min(x',x'')] } \nonumber \\
\label{eq:14adac}
\end{eqnarray}
(this simplification has only minor effect, see Ref.~\onlinecite{lip03}).
Owing to the scaling properties of the function $C(x',x'';x)$ given by Eq.\
(\ref{eq:29csu}) and of the function $u_{n}(x',x'';l)$ given by Eq.\
(\ref{eq:14adac}), the results of the calculations can be expressed essentially
in terms of three dimensionless quantities, for example, $x/S$, $\epsilon S$,
and $l/S$.  In Figs.\ \ref{fig:2}, \ref{fig:3}, and \ref{fig:4}, we show the
dependence of $\mu(x)$, $\bar{n}(x)$, $J(x)$, and $n(x)$ on $x/S$ (assuming
$x_{1} = 0$) for $\epsilon S = 1$ and various values of $l/S$ .

From Fig.\ \ref{fig:2} (lower panel), it is seen that in the diffusive limit
$l/S \rightarrow 0$ the chemical potential $\mu(x)$ decreases linearly with
$x/S$ and is continuous at the interfaces; this behavior persists in the case
of arbitrary $\epsilon S$, where $\beta [ \mu(x) - \mu_{1}] = - \epsilon x$ for
$0 \leq x/S \leq 1$.  In combination with the identical decrease of the
potential profile $E_{c}(x)$, this implies a position-independent equilibrium
density $\bar{n}(x)$ (see upper panel of Fig.\ \ref{fig:2}).  When $l/S$ rises
towards the ballistic limit $l/S \rightarrow \infty$, discontinuities of
$\mu(x)$ develop at the interfaces, which increase in magnitude, and the slope
of $\mu(x)$ becomes smaller.  This results in a rise of the equilibrium density
$\bar{n}(x)$ across the sample.  As a function of $\epsilon S$, the
discontinuities of $\mu(x)$ become smaller in magnitude if $\epsilon S
\rightarrow \infty$, and larger if $\epsilon S \rightarrow 0$, such that in the
latter case $\mu(x)$ becomes independent of $x$ for $0 < x/S < 1$.

The ratio $J(x)/J$ shown in Fig.\ \ref{fig:3} is close to unity across the
whole sample.  As the ballistic limit $l/S \rightarrow \infty$ is approached,
$J(x)/J$ becomes more and more symmetric about $x/S = 1/2$, and $J(x)/J
\rightarrow 1$ in the full range $0 \leq x/S \leq 1$.  A somewhat peculiar
behavior of $J(x)$ is observed in the diffusive regime $l/S \ll 1$.  Here,
$J(x)/J$ is very close to unity inside the sample, except for the immediate
vicinity of the ends at $x/S = 0$ and $x/S = 1$, where some structure develops,
and $J(x)/J$ converges towards a value smaller than unity when $x/S \rightarrow
0$ or $x/S \rightarrow 1$.  It is important to keep in mind the latter feature
when, within the unified treatment of spin-polarized transport, the injected
spin polarization at ferromagnet/semiconductor interfaces is defined (see Sec.\
IV.C).  When $\epsilon S$ is varied, the qualitative behavior of $J(x)/J$
persists for all values of $l/S$ considered.

In Fig.\ \ref{fig:4}, we show the ratio $n(x)/n_{0}$, where $n_{0}$ is the
constant value that $n(x)$ [as well as $\bar{n}(x)$] assume in the diffusive
limit $l/S \rightarrow 0$.  For increasing $l/S$, i.e., as the ballistic
contribution to the transport mechanism increases, the ratio $n(x)/n_{0}$
decreases as a whole.  This effect can be interpreted as reflecting the fact
that ballistically, the density decreases rapidly as the velocity rises along
the sample [see Eq.\ (\ref{eq:16uvw1})]; this is impeded at the equilibration
points, which lie very dense when $l/S \ll 1$ [slow decrease of $n(x)$] and are
widely spread when $ l/S \gg 1$ [rapid decrease of $n(x)$].  For $l/S > 1$, the
behavior of $n(x)$ is largely determined by the function $C(\epsilon x)$ [in
the ballistic limit $l/S \rightarrow \infty$, we have $n(x) = (\bar{n}/2) \, (1
+ e^{-\epsilon S}) \, C(\epsilon x)$].  Again, the qualitative behavior of
$n(x)$ does not change when $\epsilon S$ is varied.

The zero-bias limit $\epsilon \rightarrow 0$ can be treated analytically.
The solution of Eq.\ (\ref{eq:11bb}) is found to be
\begin{equation}
\chi_{1}(x) = 1 + \frac{x-x_{1}}{2l} \; .
\label{eq:11m}
\end{equation}
Since $\chi_{2}(x) = \chi_{1}(x_{0}-x)$ and $\hat{a}_{1} = \hat{a}_{2} = 1/2$
to zeroth order in $\epsilon S = \beta e V$, we obtain
\begin{equation}
\chi = 1 + \frac{S}{2l} \; , \; \; \chi_{-}(x) = \frac{x - x_{0}/2}{2l} \; .
\label{eq:11mx}
\end{equation}
From Eq.\ (\ref{eq:11aha}), the thermoballistic current $J(x)$ is found to be
constant, $J(x) = J$. For the current $J$, we have from Eqs.\ (\ref{eq:9xy})
and (\ref{eq:12stu}), to first order in $\beta e V$,
\begin{equation}
J =  \frac{2 l}{2 l + S} \, v_{e} \bar{n} \, \beta e V \; ,
\label{eq:11meeg}
\end{equation}
where $\bar{n}$ is the common value of the thermoballistic equilibrium density
at either end of the sample [see Eq.~(\ref{eq:38bbbb})]. Then, the conductance
$G = e J/V$ becomes
\begin{equation}
G = \frac{2 l}{2 l + S} \, {\cal G} \; ,
\label{eq:11heid}
\end{equation}
where ${\cal G}$ is the Sharvin interface conductance [see Eq.\ (\ref{eq:90})].
Relation (\ref{eq:11heid}) generalizes the Ohm conductance $G = (2 l/S) {\cal
G} = \sigma/S$ (valid in the diffusive regime, $l/S \ll 1$), where $\sigma = 2
\beta e^{2} v_{e} \bar{n} l$ is the conductivity.\cite{lip03} The
thermoballistic equilibrium density $\bar{n}(x)$ and the thermoballistic
density $n(x)$ are obtained from expressions (\ref{eq:12stu}) and
(\ref{eq:11ahab}), respectively, as
\begin{equation}
\bar{n}(x) = n(x) = \bar{n} \; ,
\label{eq:38cccc}
\end{equation}
i.e., they are, in the present case of zero bias, both independent of position
and equal to the equilibrium density at the ends of the sample.

\section{Spin-polarized transport within the unified description}

Having established, in the preceding section, a unified description of spinless
electron transport in semiconductors in terms of a unique thermoballistic
chemical potential, we will now extend this scheme by including the spin degree
of freedom.  We allow spin relaxation to take place during the motion of the
electrons across the ballistic intervals.  Spin relaxation is generally
governed by the equation of balance connecting the spin-polarized current with
the spin-polarized density.  In the unified description, it is the {\em
thermoballistic} current and density which enter into this equation.  The
solution of the balance equation is found in terms of a spin transport function
that is related to the spin-resolved thermoballistic chemical potentials.

\subsection{Balance equation and transport mechanism}

In a stationary situation, the total electron current $J = J_{\uparrow}(x) +
J_{\downarrow}(x)$ composed of its spin-resolved parts $J_{\uparrow}(x)$ and
$J_{\downarrow}(x)$ is conserved, whereas the {\em spin-polarized current}
$J_{-}(x) = J_{\uparrow}(x) - J_{\downarrow}(x)$, or rather its off-equilibrium
part $\hat{J}_{-}(x) = J_{-}(x) - \tilde{J}_{-}(x)$, where $\tilde{J}_{-}(x)$
is the relaxed part of the spin-polarized current, is connected with the
off-equilibrium spin-polarized density $\hat{n}_{-}(x)$ through the balance equation
\begin{equation}
\frac{d\hat{J}_{-}(x)}{dx} + \frac{\hat{n}_{-}(x)}{\tau_{s}} = 0 \; .
\label{eq:1}
\end{equation}
Here, $\hat{n}_{-}(x)$ is defined in analogy to $\hat{J}_{-}(x)$, and
$\tau_{s}$ is the spin relaxation time. For a complete description of
spin-polarized transport, the balance equation (\ref{eq:1}) is to be
supplemented with a relation between the current and the density, which
reflects the specific transport mechanism.

In the {\it ballistic limit}, the electron currents $J_{\uparrow
\downarrow}(x)$ are proportional to the densities $n_{\uparrow \downarrow}(x)$
of the electrons participating in the transport,
\begin{equation}
J_{\uparrow \downarrow}(x) = v(x) \, n_{\uparrow \downarrow}(x) \; ,
\label{eq:2}
\end{equation}
where $v(x)$ is the average velocity of the electrons at position $x$ [we
disregard spin splitting of the conduction band edge potential $E_{c}(x)$, so
that $v(x)$ is independent of spin]. This relation holds also for the
off-equilibrium spin-polarized current and density,
\begin{equation}
\hat{J}_{-}(x) = v(x) \, \hat{n}_{-}(x) \; .
\label{eq:2kaka}
\end{equation}
Use of this equation in Eq.\ (\ref{eq:1}) yields
\begin{equation}
\frac{d\hat{J}_{-}(x)}{dx} +  C(x) \, \frac{\hat{J}_{-}(x)}{l_{s}} = 0 \; ,
\label{eq:1laura}
\end{equation}
where $l_{s} = 2 v_{e} \tau_{s}$ is the (ballistic) spin relaxation length,
which comprises the overall effect of the various underlying microscopic
spin relaxation mechanisms,\cite{ell54,yaf63,dya71,yuk05} and where Eq.\
(\ref{eq:16uvw1}) has been used (omitting the positions $x', x''$ of the end
points) to express $v(x)$ in terms of $C(x)$, i.e., of the potential $E_{c}(x)$
in which the electrons move.

In the {\em diffusive regime}, the off-equilibrium spin-polarized current and
density are connected by the relation
\begin{equation}
\hat{J}_{-}(x) = - \frac{\nu}{e} \left[ \hat{n}_{-}(x) \, \frac{dE_{c}(x)}{dx}
+ \frac{1}{\beta} \, \frac{d\hat{n}_{-}(x)}{dx} \right]
\label{eq:4aa}
\end{equation}
[see Eq.\ (\ref{eq:4wasg})]. We then find from Eq.\ (\ref{eq:1})
\begin{eqnarray}
\frac{d^{2} \hat{n}_{-}(x)}{dx^{2}} &+& \beta \, \frac{dE_{c}(x)}{dx} \,
\frac{d\hat{n}_{-}(x)}{dx} \nonumber \\ &+& \beta \, \frac{d^{2}
E_{c}(x)}{dx^{2}} \, \hat{n}_{-}(x) - \frac{1}{L^{2}_{s}} \, \hat{n}_{-}(x) = 0
\; , \nonumber \\
\label{eq:5aa}
\end{eqnarray}
where
\begin{equation}
L_{s} = \sqrt{\langle l \rangle l_{s}}
\label{eq:5aaa}
\end{equation}
is the spin diffusion length.

In the {\em unified description}, the total current and density inside the
semiconducting sample are taken to be the  thermoballistic current $J(x)$ and
density $n(x)$, between which no direct relation generally exists.  Instead,
Eqs.\ (\ref{eq:11abbc}) and (\ref{eq:11abbd}) express these two quantities
separately in terms of the chemical potential $\mu(x)$.  The connection between
the off-equilibrium thermoballistic {\em spin-polarized} current
$\hat{J}_{-}(x)$ and density $\hat{n}_{-}(x)$ can be established along similar
lines, as will be described in the following.

\subsection{Thermoballistic spin-polarized current and density}

In order to include spin relaxation in the unified description, we begin by
introducing the thermoballistic equilibrium densities $\bar{n}_{\uparrow
\downarrow}(x')$ for spin-up and spin-down electrons at an equilibration point
$x'$.  It is convenient to express $\bar{n}_{\uparrow \downarrow}(x')$ in terms
of the spin-independent thermoballistic equilibrium density $\bar{n}(x')$ and a
``spin fraction'' $\alpha_{\uparrow \downarrow}(x')$ via
\begin {equation}
\bar{n}_{\uparrow \downarrow}(x') = \bar{n}(x') \, \alpha_{\uparrow
\downarrow}(x') \; ,
\label{eq:21xzzz}
\end{equation}
with $\alpha_{\uparrow}(x') + \alpha_{\downarrow}(x') = 1$.  In analogy to Eq.\
(\ref{eq:12stu}), we define spin-resolved thermoballistic chemical potentials
$\mu_{\uparrow \downarrow}(x')$ via
\begin{equation}
\bar{n}_{\uparrow \downarrow}(x') = N_{c} \, e^{-\beta [E_{c}(x') -
\mu_{\uparrow \downarrow}(x')]} \; ,
\label{eq:12nsu}
\end{equation}
which implies
\begin {equation}
e^{\beta \mu_{\uparrow \downarrow}(x')} = e^{\beta \mu(x')} \,
\alpha_{\uparrow \downarrow}(x') \; .
\label{eq:21xz}
\end{equation}
The spin fraction $\alpha_{\uparrow \downarrow}(x')$ also enters into the
definition of the spin-resolved ballistic current $J^{l}_{\uparrow
\downarrow}(x',x'')$ injected at the left end at $x'$ of the interval
$[x',x'']$,
\begin{eqnarray}
J^{l}_{\uparrow \downarrow}(x',x'') &=& v_{e} N_{c} \, e^{-\beta
[E_{c}^{m}(x',x'') - \mu_{\uparrow \downarrow}(x')]} \nonumber \\ &=&
J^{l}(x',x'') \, \alpha_{\uparrow \downarrow}(x') \; .
\label{eq:16xyz2}
\end{eqnarray}
We emphasize that expression (\ref{eq:16xyz2}) for $J^{l}_{\uparrow
\downarrow}(x',x'')$ holds only at the left end, since this current is not
conserved owing to spin relaxation, and becomes position-dependent inside the
interval.  There, we write it in the form
\begin{equation}
J^{l}_{\uparrow \downarrow}(x',x'';x) = J^{l}(x',x'') \, \alpha_{\uparrow
\downarrow}^{l}(x',x'';x) \; .
\label{eq:16x}
\end{equation}
The function $\alpha_{\uparrow \downarrow}^{l}(x',x'';x)$ is the spin fraction
at position $x$ of spin-up (spin-down) electrons injected into the ballistic
interval $[x',x'']$ at its left end at $x'$, with
$\alpha_{\uparrow}^{l}(x',x'';x) + \alpha_{\downarrow}^{l}(x',x'';x) = 1$.
Here, the dependence on the position $x''$ (at the end of the ballistic
interval $[x',x'']$ opposite to that at position $x'$ where the electrons are
injected) is due to the effect of the potential barrier embodied in the
function $E_{c}^{m}(x',x'')$ in expression (\ref{eq:16xyz1}).

When $x$ coincides with the injection point $x'$, the current (\ref{eq:16x})
becomes identical to the current (\ref{eq:16xyz2}), so that
\begin{equation}
\alpha_{\uparrow \downarrow}^{l}(x',x'';x') = \alpha_{\uparrow \downarrow}(x')
\; .
\label{eq:16bush}
\end{equation}
We now introduce the ``spin fraction excess'' $\alpha_{-}(x') =
\alpha_{\uparrow}(x') - \alpha_{\downarrow}(x')$ and the off-equilibrium spin
fraction excess $\hat{\alpha}_{-}(x') = \alpha_{-}(x') - \tilde{\alpha}_{-}$,
where $\tilde{\alpha}_{-} = \tilde{\alpha}_{\uparrow} -
\tilde{\alpha}_{\downarrow}$, and  $\tilde{\alpha}_{\uparrow \downarrow}$ are
the relaxed parts of the spin fractions ($\tilde{\alpha}_{-} = 0$ for
nonmagnetic semiconductors). With the off-equilibrium spin fraction excess
$\hat{\alpha}_{-}^{l}(x',x'';x)$ defined in an analogous way, we write the
off-equilibrium ballistic spin-polarized current $\hat{J}^{l}_{-}(x',x'';x)$ as
\begin{equation}
\hat{J}^{l}_{-}(x',x'';x) = J^{l}(x',x'') \, \hat{\alpha}_{-}^{l}(x',x'';x) \;
.
\label{eq:16xmas}
\end{equation}
The spin relaxation of the electrons injected at the left equilibration point
$x'$ into the ballistic interval $[x',x'']$ is governed by Eq.\
(\ref{eq:1laura}), which, owing to Eq.\ (\ref{eq:16x}), becomes a differential
equation for $\hat{\alpha}_{-}^{l}(x',x'';x)$,
\begin{equation}
\frac{d\hat{\alpha}_{-}^{l}(x',x'';x)}{dx} + C(x',x'';x) \,
\frac{\hat{\alpha}_{-}^{l}(x',x'';x)}{l_{s}} = 0 \; .
\label{eq:16z}
\end{equation}
The solution of Eq.\ (\ref{eq:16z}) is
\begin{equation}
\hat{\alpha}_{-}^{l}(x',x'';x) = \hat{\alpha}_{-}(x') \,
e^{-\mathfrak{C}(x',x'';x',x)/l_{s}} \; ,
\label{eq:16zz}
\end{equation}
where
\begin{equation}
\mathfrak{C}(x',x'';z_{1},z_{2}) = \int_{z_{<}}^{z_{>}} dz \,
C(x',x'';z) \; ,
\label{eq:16fdp}
\end{equation}
with $z_{<} = \min (z_{1}, z_{2})$ and  $z_{>} = \max (z_{1}, z_{2})$. We then
have for the off-equilibrium ballistic spin-polarized current at position $x$
of electrons injected at $x'$
\begin{eqnarray}
\hat{J}^{l}_{-}(x',x'';x) &=& J^{l}(x',x'') \, \hat{\alpha}_{-}(x') \,
e^{- \mathfrak{C}(x',x'';x',x)/l_{s}} \; ; \nonumber \\
\label{eq:16zzb}
\end{eqnarray}
analogously, we find
\begin{eqnarray}
\hat{J}^{r}_{-}(x',x'';x) &=& J^{r}(x',x'') \, \hat{\alpha}_{-}(x'') \,
e^{- \mathfrak{C}(x',x'';x,x'')/l_{s}} \nonumber \\
\label{eq:16zzz}
\end{eqnarray}
for the off-equilibrium ballistic spin-polarized current injected at $x''$.

Separating out the relaxed part, we now write the (net) ballistic
spin-polarized current, in analogy to Eq.\ (\ref{eq:10xx1}), in the form
\begin{equation}
J_{-}(x', x'';x) = \hat{J}_{-}(x', x'';x) + J(x',x'') \, \tilde{\alpha}_{-} \;
, \label{eq:22}
\end{equation}
with the off-equilibrium ballistic spin-polarized current
\begin{eqnarray}
&\ & \hspace*{-0.75cm} \hat{J}_{-}(x', x'';x) = v_{e} N_{c} \, e^{-\beta
\hat{E}_{c}^{m} (x',x'')} \nonumber \\ &\times& \left[ A(x') \, e^{-
\mathfrak{C}(x',x'';x',x)/l_{s}} - A(x'') \, e^{-
\mathfrak{C}(x',x'';x,x'')/l_{s}} \right] \; ; \nonumber \\
\label{eq:22a}
\end{eqnarray}
here, we have introduced the ``spin transport function''
\begin{equation}
A(x') = e^{-\beta [E_{c}^{0} - \mu(x')]} \, \hat{\alpha}_{-}(x')
\label{eq:21}
\end{equation}
at the equilibration point $x'$ ($x_{1} \leq x' \leq x_{2}$).  Using the
relation
\begin{equation}
\mu_{-}(x') = \frac{1}{\beta} \ln \left( \frac{1 + \alpha_{-}(x')}{1
- \alpha_{-}(x')} \right) \; ,
\label{eq:21ssw}
\end{equation}
between the splitting $\mu_{-}(x') = \mu_{\uparrow}(x') - \mu_{\downarrow}(x')$
of the spin-up and spin-down chemical potentials and the spin fraction excess
$\alpha_{-}(x')$, which follows from Eq.\ (\ref{eq:21xz}), we have
\begin{eqnarray}
A(x) &=& e^{-\beta [E_{c}^{0} - \mu(x)]} \nonumber \\ &\times& \left[ \tanh
\left( \frac{\beta \mu_{-}(x)}{2} \right) - \tanh \left( \frac{\beta
\bar{\mu}_{-}(x)}{2} \right) \right] \; . \nonumber \\
\label{eq:21zz}
\end{eqnarray}
For the ballistic spin-polarized density, we have, in analogy to Eq.\
(\ref{eq:22}),
\begin{equation}
n_{-}(x', x'';x) =  \hat{n}_{-}(x', x'';x) + n(x',x'';x) \, \tilde{\alpha}_{-}
\; , \label{eq:20}
\end{equation}
where
\begin{eqnarray}
&\ & \hspace*{-0.85cm} \hat{n}_{-}(x', x'';x) = \frac{N_{c}}{2} \, C(x',x'';x)
\, e^{-\beta \hat{E}_{c}^{m} (x',x'')} \nonumber \\ &\times& \left[ A(x') \,
e^{- \mathfrak{C}(x',x'';x',x)/l_{s}} + A(x'') \, e^{-
\mathfrak{C}(x',x'';x,x'')/l_{s}} \right] \nonumber \\
\label{eq:20a}
\end{eqnarray}
is the off-equilibrium ballistic spin-polarized density [see Eqs.\
(\ref{eq:10xxx1}) and (\ref{eq:22a})].

For the {\em thermoballistic spin-polarized current} $J_{-}(x)$ passing through
the point $x$, we find
\begin{equation}
J_{-}(x) = \hat{J}_{-}(x) + J(x) \, \tilde{\alpha}_{-} \; ,
\label{eq:22horst}
\end{equation}
where the off-equilibrium thermoballistic spin-polarized current
$\hat{J}_{-}(x)$ is obtained from the off-equilibrium ballistic spin-polarized
current (\ref{eq:22a}) by summing up the weighted contributions of the
ballistic intervals,
\begin{eqnarray}
&\ & \hspace*{-0.8cm} \hat{J}_{-}(x) = v_{e} N_{c} \, \int_{x_{1}}^{x}
\frac{dx'}{l} \int_{x}^{x_2} \frac{dx''}{l} \, \mathbb{W}(x',x'';l)
\nonumber \\ &\times& \left[ A(x') \, e^{- \mathfrak{C}(x',x'';x',x)/l_{s}}  -
A(x'') \, e^{- \mathfrak{C}(x',x'';x,x'')/l_{s}} \right] \nonumber \\
\label{eq:23}
\end{eqnarray}
$(x_{1} < x < x_{2})$. Similarly, the {\em thermoballistic spin-polarized
density} $n_{-}(x)$ at the point $x$ is
\begin{equation}
n_{-}(x) = \hat{n}_{-}(x) + n(x) \, \tilde{\alpha}_{-} \; ,
\label{eq:22eva}
\end{equation}
where the off-equilibrium thermoballistic spin-polarized density
$\hat{n}_{-}(x)$ is obtained from (\ref{eq:20a}) as
\begin{eqnarray}
&\ & \hspace*{-0.85cm} \hat{n}_{-}(x) =  \frac{N_{c}}{2} \, \int_{x_{1}}^{x}
\frac{dx'}{l} \int_{x}^{x_2} \frac{dx''}{l} \, \mathbb{W}_{C}(x',x'';l;x)
\nonumber \\ &\times& \left[ A(x') \, e^{- \mathfrak{C}(x',x'';x',x)/l_{s}} +
A(x'') \, e^{- \mathfrak{C}(x',x'';x,x'')/l_{s}} \right] \nonumber \\
\label{eq:24}
\end{eqnarray}
$(x_{1} < x < x_{2})$.

In the diffusive regime, $l/l_{s} \ll 1$, $l/S \ll 1$, the integrals over $x'$
and $x''$ in Eqs.\ (\ref{eq:23}) and (\ref{eq:24}) can be evaluated explicitly,
yielding
\begin{equation}
\hat{J}_{-}(x) = - 2 v_{e} N_{c} l \, e^{-\beta [E_{c}(x) - E_{c}^{0}]} \,
\frac{d A(x)}{dx}
\label{eq:23x}
\end{equation}
and
\begin{equation}
\hat{n}_{-}(x) = N_{c} \, e^{-\beta [E_{c}(x) - E_{c}^{0}]} \, A(x) \; .
\label{eq:24x}
\end{equation}
Eliminating the function $A(x)$ from these two equations, we obtain the
standard drift-diffusion relation (\ref{eq:4aa}).

In the current $\hat{J}_{-}(x)$ and density $\hat{n}_{-}(x)$, the spin
relaxation in each ballistic interval is described in terms of the values
of the spin transport function $A(x)$ at the end points $x = x'$ and $x = x''$.
Since $\hat{J}_{-}(x)$ and $\hat{n}_{-}(x)$ are {\em linearly} connected with
$A(x)$, they are linearly connected with each other.  Thus, it appears that the
function $A(x)$ [and not the chemical-potential splitting $\mu_{-}(x)$] is the
key quantity for treating spin transport within the unified description.  It
remains to find an equation for the determination of this function.

\subsection{Integral equation for the spin transport function}

The required equation is provided by the basic balance equation, Eq.\
(\ref{eq:1}), which we now read in terms of the off-equilibrium thermoballistic
spin-polarized current (\ref{eq:23}) and density (\ref{eq:24}) of the unified
transport description.  Since the derivative with respect to $x$ of the terms
in the brackets of expression (\ref{eq:23}) for $\hat{J}_{-}(x)$ is compensated
by the term $\hat{n}_{-}(x)/\tau_{s}$ (this reflects the fact that spin
relaxation in the ballistic intervals has already been taken into account),
only the derivative on the limits of integration in expression (\ref{eq:23})
remains, and we have
\begin{eqnarray}
&\ & \hspace*{-0.8cm} \int_{x}^{x_{2}} \frac{dx'}{l} \,
\mathbb{W}(x,x';l) \left[ A(x) - A(x') \, e^{- \mathfrak{C}(x,x')/l_{s}}
\right] \nonumber \\ &-& \int_{x_{1}}^{x} \frac{dx'}{l} \, \mathbb{W}(x',x;l)
\left[ A(x') \, e^{- \mathfrak{C}(x',x)/l_{s}} - A(x) \right] = 0 \; ,
\nonumber \\
\label{eq:26}
\end{eqnarray}
where
\begin{equation}
\mathfrak{C}(x',x'') = \mathfrak{C}(x',x'';x',x'') \; .
\label{eq:16kpd}
\end{equation}
With the action of the symbolic operator $\mathbb{W}(x',x'';l)$ explained  by
comparison of Eqs.\ (\ref{eq:11a}) and (\ref{eq:11aba}), Eq.~(\ref{eq:26})
reads explicitly
\begin{eqnarray}
{\cal W}_{2}(x_{1},x;l,l_{s}) \, A_{1} &+& {\cal W}_{2} (x,x_{2};l,l_{s})
\, A_{2} \nonumber \\ &-& W(x_{1},x_{2};x;l) \, A(x)  \nonumber \\ &+&
\int_{x_{1}}^{x_{2}} \frac{dx'}{l} \, {\cal W}_{3}(x',x;l,l_{s})
\, A(x') = 0 \; , \nonumber \\
\label{eq:26a}
\end{eqnarray}
where
\begin{equation}
{\cal W}_{n}(x',x'';l,l_{s}) = w_{n}(x',x'';l) \, e^{- \mathfrak{C}
(x',x'')/l_{s}} \; ,
\label{eq:26abc}
\end{equation}
\begin{eqnarray}
W(x_{1},x_{2};x;l) &=& w_{2}(x_{1},x;l) + w_{2}(x,x_{2};l) \nonumber \\ &+&
\int_{x_{1}}^{x_{2}} \frac{dx'}{l} w_{3}(x',x;l) \; ,
\label{eq:26bcd}
\end{eqnarray}
and $A_{1,2} = A(x_{1,2})$.  Equation (\ref{eq:26a}) is a linear, Fredholm-type
integral equation for the spin transport function $A(x)$.  Its solution for
$x_{1} < x < x_{2}$ determines the spin-polarized electron transport inside the
semiconducting sample, and is obtained in terms of the values $A_{1}$ and
$A_{2}$ at the interfaces at the ends of the sample.  The latter are determined
by the off-equilibrium spin fraction excesses $\hat{\alpha}_{-}(x_{1,2})$ and
the chemical potentials $\mu_{1,2} = \mu(x_{1,2})$ in the contacts [see Eq.\
(\ref{eq:21})].  The function $A(x)$ is not, in general, continuous at the
interfaces, $A(x_{1}^{+}) \neq A_{1}$, $A(x_{2}^{-}) \neq A_{2}$ (``Sharvin
effect''), as will be demonstrated in Sec.\ III.D by way of a particular
example.  The discontinuities of $A(x)$ arise from the joint effect of the
discontinuities of the spin-independent thermoballistic chemical potential
$\mu(x)$ and those of the spin fraction excess $\alpha_{-}(x)$ [or,
equivalently, of the spin-resolved thermoballistic chemical potentials
$\mu_{\uparrow \downarrow}(x)$].

Substituting $A(x)$ in Eqs.\ (\ref{eq:23}) and (\ref{eq:24}), we obtain the
off-equilibrium thermoballistic spin-polarized current $\hat{J}_{-}(x)$ and
density $\hat{n}_{-}(x)$, respectively;  the thermoballistic spin-polarized
current $J_{-}(x)$ and density $n_{-}(x)$ then follow from Eqs.\
(\ref{eq:22horst}) and (\ref{eq:22eva}), respectively.  Dividing by the
corresponding total thermoballistic current and density, Eqs.\
(\ref{eq:11abbc}) and (\ref{eq:11abbd}), respectively, we get the current spin
polarization
\begin{equation}
P_{J}(x) = \frac{J_{-}(x)}{J(x)} = \frac{\hat{J}_{-}(x)}{J(x)} +
\tilde{\alpha}_{-}
\label{eq:40a}
\end{equation}
and the density spin polarization
\begin{equation}
P_{n}(x) = \frac{n_{-}(x)}{n(x)} = \frac{\hat{n}_{-}(x)}{n(x)} +
\tilde{\alpha}_{-}
\label{eq:40b}
\end{equation}
inside the sample.  These polarizations are written in terms of the
thermoballistic current and density; however, we take their magnitudes to
be also those of the {\em physical polarizations}, for the following reason.
The underlying assumption of our approach is that the equilibration process,
i.e., the coupling between the thermoballistic and background currents, is
independent of spin (this is clearly true for the D'yakonov-Perel' spin
relaxation mechanism,\cite{dya71} but remains to be examined for the other
mechanisms).  Therefore, the relative spin content is the same in these two
currents, and thus equal to that of their sum, {\em viz.}, the physical
current.  Hence, we may take the polarizations $P_{J}(x)$ and $P_{n}(x)$ of
Eqs.\ (\ref{eq:40a}) and (\ref{eq:40b}), respectively, for the physical
polarizations.

The integral equation (\ref{eq:26a}) constitutes the central result of the
present work.  It allows the calculation of the spin polarization in
semiconductors for any value of the momentum and spin relaxation lengths as
well as for arbitrary band edge potential profile.  The fact that we are led,
in the unified description of spin-polarized transport, to an integral equation
is connected with the introduction of the momentum relaxation length $l$ as an
independent parameter of arbitrary magnitude, which gives rise to nonlocal
ballistic effects.  The basic parameters controlling the transport in the
unified description are the equilibrium densities $\bar{n}_{1,2}$, the momentum
relaxation length $l$, and the spin relaxation length $l_{s}$, whereas in the
standard drift-diffusion model one uses the conductivity $\sigma$ and the spin
diffusion length $L_{s}$.

\subsection{Differential equation for the spin transport function}

In order to interpret our unified description of spin-polarized transport and
relate it to previous, less general descriptions, we consider in the following
the case of field-driven transport in a homogeneous semiconductor without
space charge.  As in Sec.\ II.C, we take the probabilities $p_{n}(x/l)$ in
their one-dimensional form.  Then, Eq.\ (\ref{eq:26a}) can be converted into an
integrodifferential equation for the spin transport function $A(x)$.  In an
approximation which is adequate for the present purposes, the latter equation
reduces to a second-order differential equation.

\subsubsection{General form and diffusive regime}

For a potential of the form (\ref{eq:38aaaa}), the integral equation
(\ref{eq:26a}) reduces, with the help of Eqs.\ (\ref{eq:29csu}),
(\ref{eq:14adac}), (\ref{eq:16fdp}), and (\ref{eq:16kpd}), to
\begin{eqnarray}
&\ & \hspace*{-1.45cm} f_{1}(x-x_{1}) \, A_{1} +f_{2}(x_{2}-x) \, A_{2}
\nonumber \\ &-& f(x-x_{1}) \, A(x)  + \int_{x_{1}}^{x} \frac{dx'}{l} \,
f_{1}(x-x') \, A(x')  \nonumber \\ &+& \int_{x}^{x_{2}} \frac{dx'}{l} f_{2}(x'-
x) \, A(x') = 0 \; ,
\label{eq:38b}
\end{eqnarray}
where
\begin{equation}
f_{1}(x) =  e^{-[\epsilon + 1/l + \mathfrak{c} (\epsilon x)/l_{s}] x} \; ,
\label{eq:38ijk}
\end{equation}
\begin{equation}
f_{2}(x) =  e^{-[1/l + \mathfrak{c} (\epsilon x)/l_{s}] x} \; ,
\label{eq:38jkl}
\end{equation}
\begin{equation}
f(x) = \frac{1}{1 + \epsilon l} \left\{ 2 + \epsilon l \left[ 1 +
e^{-(\epsilon + 1/l)x} \right] \right\} \; ,
\label{eq:38klm}
\end{equation}
and
\begin{equation}
\mathfrak{c}(\zeta) = \frac{1}{\zeta} \int_{0}^{\zeta} d \zeta' \,
C(\zeta')
\label{eq:38pds}
\end{equation}
with $0 < \mathfrak{c}(\zeta) \leq 1$, $\mathfrak{c}(\zeta) \rightarrow1 $ for
$\zeta \rightarrow 0$, and $\mathfrak{c}(\zeta) \sim 2 (\pi \zeta)^{-1/2}$ for
$\zeta \rightarrow \infty$.

By supplementing the inhomogeneous integral equation (\ref{eq:38b}) with the
equations obtained by forming its first and second derivative with respect to
$x$, and eliminating from this set of equations the quantities $A_{1}$ and
$A_{2}$, we can convert Eq.\ (\ref{eq:38b}) into a homogeneous
integrodifferential equation for the spin transport function $A(x)$.  [This
procedure could also be applied to Eq.\ (\ref{eq:26a}), but does not seem to be
helpful in the general case].  Now, the latter equation can be simplified by
replacing the function $\mathfrak{c}(\zeta)$ with a position-independent
average value $\bar{\mathfrak{c}}$, so that the coefficient functions
$f_{1}(x)$ and $f_{2}(x)$ in Eq.~(\ref{eq:38b}) reduce to pure exponentials.
With this approximation, the integrodifferential equation for $A(x)$ becomes a
second-order differential equation of the form
\begin{equation}
b_{0}(x) \, \frac{d^{2} A(x)}{dx^{2}} + b_{1}(x) \, \frac{d A(x)}{dx} +
b_{2}(x) \, A(x)  = 0 \; ,
\label{eq:38csu}
\end{equation}
where
\begin{equation}
b_{0}(x) = 2 + \epsilon l [ 1 + b(x)] \; ,
\label{eq:38ca}
\end{equation}
\begin{equation}
b_{1}(x) =  \epsilon (2 + \epsilon l) [ 1 - b(x)] \; ,
\label{eq:38cb}
\end{equation}
\begin{eqnarray}
b_{2}(x) &=& \frac{1}{l \bar{l} {\tilde{l}}^{2}} \left\{ 2 \left[
{\tilde{l}}^{2} - l \bar{l} +\epsilon l \tilde{l} (\tilde{l} - \bar{l}) \right]
\right.  \nonumber \\ &+& \left.  \rule{0mm}{4mm}\epsilon \bar{l}(\tilde{l} -
l) (l + \tilde{l} + \epsilon l \tilde{l}) \left[ 1 + b(x)\right] \right\} \; ,
\label{eq:38cc}
\end{eqnarray}
with
\begin{equation}
b(x) = e^{-(\epsilon + 1/l)(x-x_{1})}
\label{eq:38caxx}
\end{equation}
and
\begin{equation}
\frac{1}{\bar{l}} =\frac{1}{l} + \frac{1}{l_{s}} \; , \; \;
\frac{1}{\tilde{l}} =\frac{1}{l} + \frac{1}{\tilde{l}_{s}}
\; , \; \; \tilde{l}_{s} = \frac{l_{s}}{\bar{\mathfrak{c}}} \; .
\label{eq:38xxac}
\end{equation}
Since, owing to the presence of the factor $e^{-x/l}$ in the functions
$f_{1}(x)$ and $f_{2}(x)$, only the values of $\mathfrak{c}(\epsilon x)$ within
the range $0 \leq x \alt l$ contribute appreciably, we choose
$\bar{\mathfrak{c}}$ as the average of $\mathfrak{c}(\epsilon x)$ over an
$x$-interval of length equal to the momentum relaxation length $l$,
\begin{equation}
\bar{\mathfrak{c}}  = \frac{1}{l} \int_{0}^{l} dx \, \mathfrak{c}(\epsilon
x) =  \frac{1}{\epsilon l} \int_{0}^{\epsilon l} d \zeta \, \ln ( \epsilon
l/\zeta) \, C(\zeta)   \; .
\label{eq:38tux}
\end{equation}
In the right-hand integral of this equation, the range of small $\epsilon x$,
where $C(\epsilon x) \approx 1$, is emphasized because of the weight factor $
\ln ( \epsilon l/\zeta)$.  For large $\epsilon l$ (in the ballistic regime
and/or for strong fields), the variation of $\mathfrak{c}(\epsilon x)$ with $x$
becomes essential, and a more detailed study of the validity of the
approximation leading to Eq.\ (\ref{eq:38csu}) will be necessary.  For the
present purpose of solely demonstrating the principal effects of the transport
mechanism, we consider this approximation, in conjunction with the choice
(\ref{eq:38tux}) for $\bar{\mathfrak{c}}$, to be sufficiently accurate.

In the diffusive regime characterized by the conditions $l/l_{s} \ll 1$, $l/S
\ll 1$, and $\epsilon l \ll 1$, we have $\bar{\mathfrak{c}} = 1$ and $\tilde{l}
= \bar{l}$, and Eq.\ (\ref{eq:38csu}) reduces to
\begin{equation} \frac{d^{2} A(x)}{dx^{2}} + \epsilon \frac{d A(x)}{dx} -
\frac{1}{L_{s}^{2}} A(x) = 0 \; .
\label{eq:38d}
\end{equation}
In view of Eq.\ (\ref{eq:24x}), Eq.\ (\ref{eq:38d}) can be rewritten in terms
of $\hat{n}_{-}(x)$ and then agrees with Eq.~(\ref{eq:5aa}), and with Eq.~(2.8)
of Yu and Flatt\'{e}\cite{yuf02a} if the intrinsic spin diffusion length $L$ of
that reference is identified with the spin diffusion length $L_{s} = \sqrt{l
l_{s}}$.  It thus turns out that Eq.\ (\ref{eq:38csu}) generalizes the usual
spin drift-diffusion equation to the case of arbitrary values of the ratio
$l/l_{s}$.

\subsubsection{Zero-bias limit}

In the zero-bias limit $\epsilon \rightarrow 0$, the integral equation
(\ref{eq:38b}) reduces to
\begin{eqnarray}
&\ & \hspace*{-1.75cm} e^{-(x-x_{1})/\bar{l}} A_{1} + e^{-(x_{2}-x)/\bar{l}}
A_{2} \nonumber \\ &-& 2 A(x)+ \int_{x_{1}}^{x_{2}} \frac{dx'}{l} e^{-|x-
x'|/\bar{l}}A(x') = 0 \;
,
\label{eq:28}
\end{eqnarray}
from which one derives the differential equation
\begin{equation}
\frac{d^{2}A(x)}{dx^{2}} - \frac{1}{L^{2}} A(x) = 0 \; ;
\label{eq:30}
\end{equation}
here,
\begin{equation}
L = \sqrt{\bar{l} l_{s}} = \frac{L_{s}}{\sqrt{1 + l/l_{s}}}
\label{eq:31}
\end{equation}
is the generalization of the spin diffusion length (\ref{eq:5aaa}), which
includes ballistic effects via the renormalization factor $1/\sqrt{1 +
l/l_{s}}$.  The length $L$ becomes equal to the spin diffusion length proper,
$L = L_{s}$, in the diffusive regime where $\bar{l} = l $, and to the spin
relaxation length, $L = l_{s}$, in the ballistic limit $l/l_{s}
\rightarrow \infty$ where $\bar{l} = l_{s}$.

Equation (\ref{eq:30}) has the general solution, for $x_{1}< x < x_{2}$,
\begin{equation}
A(x) = C_{1} e^{-(x-x_{1})/L} + C_{2} e^{-(x_{2}-x)/L} \; .
\label{eq:32}
\end{equation}
With this expression substituted  for $A(x)$ in Eq.\ (\ref{eq:28}), the set of
two equations resulting from writing down this equation for $x = x_{1}$ and $x
= x_{2}$, respectively, can be solved for the  coefficients  $C_{1,2}$,
\begin{equation}
C_{1} = \frac{1}{D} \left[ (1+\gamma) e^{S/L} A_{1} - (1-\gamma) A_{2}\right]
\; ,
\label{eq:33}
\end{equation}
\begin{equation}
C_{2} = - \frac{1}{D} \left[ (1-\gamma) A_{1}- (1+\gamma) e^{S/L} A_{2} \right]
\; ,
\label{eq:33abc}
\end{equation}
where
\begin{equation}
D =(1+\gamma)^{2} e^{S/L} -(1-\gamma)^{2} e^{-S/L} \; ,
\label{eq:34}
\end{equation}
with
\begin{equation}
\gamma = \frac{L}{l_{s}} = \frac{\bar{l}}{L} = \sqrt{\frac{l}{l+l_{s}}} \leq
1 \; .
\label{eq:35}
\end{equation}
It follows from Eqs.\ (\ref{eq:32})--(\ref{eq:35}) that the function
$A(x)$ is discontinuous at $x = x_{1,2}$,
\begin{equation}
\Delta A_{1} \equiv A(x_{1}^{+}) - A_{1} = - \frac{1}{2} \, (g A_{1} - h A_{2})
\; ,
\label{eq:35aha}
\end{equation}
\begin{equation}
\Delta A_{2} \equiv A_{2} - A(x_{2}^{-}) = - \frac{1}{2} \, (h A_{1}
- g A_{2}) \; ,
\label{eq:35ahax}
\end{equation}
where
\begin{equation}
g = \frac{2 \gamma}{D} \left[ (1+\gamma)\, e^{S/L} + (1-\gamma) \,e^{-S/L}
\right] \leq 1
\label{eq:61b}
\end{equation}
and
\begin{equation}
h = \frac{4 \gamma}{D} \leq \frac{2}{1 + \gamma} \, e^{-S/L} \; .
\label{eq:61bibi}
\end{equation}
In the diffusive regime, one has $L = L_{s} = l_{s} \sqrt{l/l_{s}}$ and $\gamma
= 0$, and therefore
\begin{equation}
A(x) = A_{1} \, e^{- (x-x_{1})/L_{s}} +  A_{2} \, e^{-(x_{2}-x)/L_{s}} \; .
\label{eq:37}
\end{equation}
In the ballistic limit, one has $L = l_{s}$ and $\gamma = 1$, so that
\begin{equation}
A(x) = \frac{1}{2} \left[ A_{1} \, e^{-(x-x_{1})/l_{s}} + A_{2} \,
e^{-(x_{2}-x)/l_{s}} \right] \;  .
\label{eq:36}
\end{equation}
The discontinuity of $A(x)$, e.g., at $x = x_{1}$, is $\Delta A_{1} = A_{2}
\exp(-S/L_{s})$ in the diffusive regime, and $\Delta A_{1} = \smfrac{1}{2}[ -
A_{1} + A_{2} \exp(-S/l_{s})]$ in the ballistic limit.

\section{Spin-polarized transport in ferromagnet/semiconductor
heterostructures}

We now turn to the unified description of spin-polarized electron transport in
heterostructures formed of a semiconductor and two ferromagnetic contacts (cf.\
Fig.\ \ref{fig:1}).  We treat the ferromagnets as fully degenerate Fermi
systems.  The semiconductor is taken to be nonmagnetic [i.e.,
$\tilde{\alpha}_{-} = 0$, and hence $\hat{\alpha}_{-}(x') = \alpha_{-}(x')$]
and homogeneous without space charge.  We disregard spin-flip scattering at
the interfaces, but spin-selective interface resistances are included in our
description by introducing discontinuities into the spin-resolved chemical
potentials, in the same way as in previous
descriptions\cite{smi01,fer01,ras02a,yuf02a} within the drift-diffusion model.
Of course, for realistic applications, it is necessary to treat the effect of
interface barriers explicitly,\cite{han03} using potential profiles $E_{c}(x)$
which, generally, must be calculated self-consistently from a nonlinear Poisson
equation.  In some cases, however, it may be sufficient to perform
non-self-consistent calculations using appropriately modeled band edge
profiles.\cite{alb02,alb03,she04} In any case, this would require the spin
transport function $A(x)$ to be determined by numerically solving the integral
equation (\ref{eq:26a}).  This task will be deferred to future work.

In order to obtain the position dependence of the spin polarization across the
heterostructure, the current spin polarization and the chemical potential in
the semiconductor are to be connected with the corresponding quantities in the
left and right ferromagnets.  The current spin polarization $P_{J}(x)$ [see
Eq.\ (\ref{eq:40a})], as expressed by the ratio of the thermoballistic currents
$\hat{J}_{-}(x)$ and $J(x)$, is equal to the physical current spin polarization
[see the discussion following Eq.\ (\ref{eq:40b})].  It is, therefore,
continuous across the whole heterostructure, in particular, at the interfaces,
so that the current spin polarizations in the semiconductor and the
ferromagnets can be equated directly there.  On the other hand, in the presence
of ballistic contributions, the thermoballistic chemical potential $\mu(x)$ and
the spin transport function $A(x)$ are {\em not} continuous at the interfaces.
The discontinuities at the interfaces are taken into account when the functions
$\mu(x)$ and $A(x)$ inside the semiconductor are calculated in terms of their
values $\mu_{1,2}$ and $A_{1,2}$, respectively.  The latter values are to be
equated with the values of the corresponding quantities in the ferromagnet.

We begin with a brief summary of the standard description (see, e.g.,
Ref.~\onlinecite{yuf02a}) of the spin polarization in the ferromagnets.

\subsection{Current spin polarization in the ferromagnets}

In the (semi-infinite) left ferromagnet located in the range $x < x_{1}$,
the spin-up and spin-down chemical potentials $\mu_{\uparrow \downarrow}(x)$
are given by
\begin{equation}
\mu_{\uparrow \downarrow}(x) = \frac{e^{2} J}{\sigma_{1}} \, (x_{1} - x)
\pm \frac{C_{1}} {\sigma_{\uparrow \downarrow}^{(1)}} \,
e^{-(x_{1}-x)/L_{s}^{(1)}} \; ,
\label{eq:50}
\end{equation}
where $L_{s}^{(1)}$ is the spin diffusion length.  The quantities
$\sigma_{\uparrow \downarrow}^{(1)}$ are the conductivities for spin up and
spin down, which are independent of position, and $\sigma_{1} =
\sigma_{\uparrow}^{(1)} + \sigma_{\downarrow}^{(1)}$. We then have
$C_{1} = \sigma_{\uparrow}^{(1)} \, \mu_{\uparrow}(x_{1}^{-}) = -
\sigma_{\downarrow}^{(1)} \mu_{\downarrow}(x_{1}^{-})$, and therefore $C_{1} =
(\sigma_{\uparrow}^{(1)} \sigma_{\downarrow}^{(1)}/\sigma_{1}) \,
\mu_{-}(x_{1}^{-})$, where $\mu_{-}(x) = \mu_{\uparrow}(x) -
\mu_{\downarrow}(x)$. With
\begin{equation}
J_{\uparrow \downarrow}(x) =  - \frac{\sigma_{\uparrow
\downarrow}^{(1)}}{e^{2}} \, \frac{d \mu_{\uparrow \downarrow}(x)}{dx} \; ,
\label{eq:56}
\end{equation}
we now find for the current spin polarization
\begin{equation}
P_{J}(x) = P_{1} - \frac{G_{1}}{2 e^{2}J} \, \mu_{-}(x_{1}^{-}) \,
e^{-(x_{1}-x)/L_{s}^{(1)}} \, ,
\label{eq:57}
\end{equation}
where $P_{1} = (\sigma_{\uparrow}^{(1)} - \sigma_{\downarrow}^{(1)})/
\sigma_{1}$ is the relaxed (current or density) spin polarization in the left
ferromagnet, and
\begin{equation}
G_{1} = \frac{4 \sigma_{\uparrow}^{(1)} \sigma_{\downarrow}^{(1)}}{\sigma_{1}
L_{s}^{(1)}} = \frac{\sigma_{1}}{L_{s}^{(1)}} \, \left( 1 - P_{1}^{2} \right)
\label{eq:57xyz}
\end{equation}
is a transport parameter of the ferromagnet, which has the dimension of
interface conductance.  Analogously, we obtain
\begin{equation}
P_{J}(x) = P_{2}  + \frac{G_{2}}{2 e^{2} J} \, \mu_{-}(x_{2}^{+}) \,
e^{-(x-x_{2})/L_{s}^{(2)}}
\label{eq:57zzz}
\end{equation}
for the current spin polarization in the right ferromagnet located in the range
$x > x_{2}$.

In the absence of spin-selective interface resistances, the chemical-potential
splitting $\mu_{-}(x)$ is continuous at the interface, $\mu_{-}(x_{1}^{-}) =
\mu_{-}(x_{1})$ and $\mu_{-}(x_{2}) = \mu_{-}(x_{2}^{+})$, where
$\mu_{-}(x_{1,2})$ are its values at the interface itself.  The latter are to
be set equal to the corresponding values in the semiconductor, which yields
\begin{eqnarray}
[\mu_{-}(x_{1,2})]_{\rm ferromagnet} &=& [\mu_{-}(x_{1,2})]_{\rm semiconductor}
\nonumber \\ &=& \frac{1}{\beta} \ln \left( \frac{1 + \alpha_{1,2}}{1
-\alpha_{1,2}} \right) \; ,
\label{eq:68}
\end{eqnarray}
where the right-hand part of this equation follows from Eq.\ (\ref{eq:21ssw})
for $x' = x_{1,2}$, and $\alpha_{1,2} = \alpha_{-}(x_{1,2})$. For the current
spin polarizations at the interfaces, $P_{J}(x_{1,2})$, we have from Eqs.\
(\ref{eq:57}) and (\ref{eq:57zzz})
\begin{equation}
\hspace*{-0.6cm} P_{J}(x_{1}) = P_{1} - \frac{G_{1}}{2 \beta e^{2} J} \, \ln
\left( \frac{1 + \alpha_{1}}{1 -\alpha_{1}} \right)  \, ,
\label{eq:57angi}
\end{equation}
\begin{equation}
\hspace*{0.3cm} P_{J}(x_{2}) = P_{2} + \frac{G_{2}}{2  \beta e^{2} J} \, \ln
\left( \frac{1 + \alpha_{2}}{1 -\alpha_{2}} \right)  \, ,
\label{eq:57guido}
\end{equation}
which are to be set equal to the corresponding polarizations of the
semiconductor.

Spin-selective interface resistances $\rho_{\uparrow \downarrow}^{(1,2)}$
are introduced via discontinuities of the spin-resolved chemical potentials
on the contact sides of the interfaces.   At $x = x_{1}$, for example, the
discontinuity has the form
\begin{equation}
\mu_{\uparrow \downarrow}(x_{1}^{-}) - \mu_{\uparrow \downarrow}(x_{1}) = e^{2}
J_{\uparrow \downarrow}(x_{1}) \, \rho_{\uparrow \downarrow}^{(1)} \; .
\label{eq:ratz}
\end{equation}
The corresponding interface resistance is located between $x = x_{1}^{-}$ and
$x_{1}$ (in the ferromagnetic contact), and thus is adjacent to the Sharvin
interface resistance between $x = x_{1}$ and $x_{1}^{+}$ (in the
semiconductor).  The quantity $\mu_{-}(x_{1}^{-})$ to be substituted in Eq.\
(\ref{eq:57}) is obtained, using Eqs.\ (\ref{eq:50}), (\ref{eq:56}), and
(\ref{eq:57xyz}), as
\begin{eqnarray}
\mu_{-}(x_{1}^{-}) &=& \frac{1}{1 + G_{1} \rho_{+}^{(1)}/4 } \nonumber \\ &\ &
\times \left\{ \mu_{-}(x_{1}) + \frac{e^{2} J}{2} \left[ P_{1} \,
\rho_{+}^{(1)} + \rho_{-}^{(1)} \right] \right\} \; , \nonumber \\
\label{eq:karol}
\end{eqnarray}
where $\rho_{\pm}^{(1)} = \rho_{\uparrow}^{(1)} \pm \rho_{\downarrow}^{(1)}$.
The connection of $\mu_{-}(x_{1})$ with the spin fraction excess $\alpha_{1}$
is, as before, given by Eq.\ (\ref{eq:68}).  The same procedure applies {\em
mutatis mutandis} to the interface at $x = x_{2}$.

\subsection{Spin polarization across a heterostructure in the zero-bias limit}

In the zero-bias limit $J \rightarrow 0$, we now demonstrate the procedure for
calculating the current and density spin polarizations across a
ferromagnet/semiconductor heterostructure.

Evaluating expressions (\ref{eq:23}) and (\ref{eq:24}), respectively, with
$A(x)$ given by Eq.\ (\ref{eq:32}), we find for the thermoballistic
spin-polarized current in the semiconductor
\begin{equation}
J_{-}(x) = - 2 v_{e} N_{c} \bar{l} \,  \frac{d A(x)}{dx}  \; ,
\label{eq:40}
\end{equation}
and for the thermoballistic spin-polarized density
\begin{equation}
n_{-}(x)  = N_{c} \, A(x) \label{eq:39}
\end{equation}
$(x_{1} < x < x_{2})$. For zero bias, one has $J(x) = J = {\rm const.}$ and
$n(x) = \bar{n} = {\rm const.}$ [see Eqs.\ (\ref{eq:11meeg}) and
(\ref{eq:38cccc})], so that, by combining expressions (\ref{eq:40}) and
(\ref{eq:39}), we obtain the relation
\begin{equation}
P_{J}(x) = - \frac{2 v_{e} \bar{n} \bar{l}}{J}  \,  \frac{d P_{n}(x)}{dx}
\label{eq:40frey}
\end{equation}
between the current and density spin polarizations.  Furthermore, $J_{-}(x)$
and $n_{-}(x)$ both satisfy Eq.\ (\ref{eq:30}), and so do the polarizations
$P_{J}(x)$ and $P_{n}(x)$ given by Eqs.\ (\ref{eq:40a}) and (\ref{eq:40b}),
respectively. Differentiation of Eq.~(\ref{eq:40frey}) then yields, together
with Eq.\ (\ref{eq:30}) for $P_{n}(x)$,
\begin{equation}
P_{n}(x) = - \frac{l_{s} J}{2 v_{e} \bar{n}}  \,  \frac{d P_{J}(x)}{dx} \; .
\label{eq:40rice}
\end{equation}
From Eq.\ (\ref{eq:40a}) with $J(x) = J$, we find, using Eqs.\ (\ref{eq:32})
and (\ref{eq:40}), the explicit form of the current spin polarization as
\begin{eqnarray}
P_{J}(x) &=&  \frac{2 v_{e} N_{c} \bar{l}}{LJ} \, \left[ C_{1} \,
e^{-(x-x_{1})/L} - C_{2} \, e^{-(x_{2}-x)/L} \right]  \; . \nonumber \\
\label{eq:40c}
\end{eqnarray}
The density spin polarization is obtained from Eqs.\ (\ref{eq:32}) and
(\ref{eq:39}) [or, equivalently, from Eqs.\ (\ref{eq:40rice}) and
(\ref{eq:40c})] as
\begin{equation}
P_{n}(x) =  \frac{N_{c}}{\bar{n}} \, \left[ C_{1} \, e^{-(x-x_{1})/L} + C_{2}
\, e^{-(x_{2}-x)/L} \right]  \; .
\label{eq:40tra}
\end{equation}
The coefficients $C_{1,2}$ in Eqs.\ (\ref{eq:40c}) and (\ref{eq:40tra}) can be
expressed via Eqs.\ (\ref{eq:33})--(\ref{eq:35}), using Eq.\ (\ref{eq:21}), in
terms of the spin fraction excesses $\alpha_{1,2}$ on the contact sides of the
interfaces.

In order to determine the quantities $\alpha_{1,2}$, we consider the current
spin polarization  (\ref{eq:40c}) on the semiconductor sides of the
interfaces,
\begin{equation}
P_{J}(x_{1}^{+}) = \frac{{\cal G}}{\beta e^{2}J} \, (g \alpha_{1} - h
\alpha_{2}) \; , \label{eq:61a}
\end{equation}
\begin{equation}
P_{J}(x_{2}^{-}) = \frac{{\cal G}}{\beta e^{2}J} \, (h \alpha_{1}
- g \alpha_{2}) \; ;
\label{eq:61abc}
\end{equation}
here, ${\cal G}$ is the Sharvin interface conductance given by Eq.\
(\ref{eq:90}), and the coefficients $g$ and $h$ are given by Eqs.\
(\ref{eq:61b}) and (\ref{eq:61bibi}), respectively.  As mentioned before,
$P_{J}(x_{1}) = P_{J}(x_{1}^{+})$ and $P_{J}(x_{2}^{-}) = P_{J}(x_{2})$, and
the connection with the polarization in the contacts is made by equating
expressions (\ref{eq:57angi}) and (\ref{eq:61a}), and expressions
(\ref{eq:57guido}) and (\ref{eq:61abc}) [note that, if spin-selective interface
resistances are included, expression (\ref{eq:57angi}) for $P_{J}(x_{1})$ is to
be replaced with the general expression obtained by using expression
(\ref{eq:karol}) for $\mu_{-}(x_{1}^{-})$ in Eq.\ (\ref{eq:57}), and
analogously for $P_{J}(x_{2})$]. In the zero-bias limit, when $J(x_{1,2}) = J$
[or $\kappa = 1$ in Eq.\ (\ref{eq:5e})], we have $|\alpha_{1,2}| \ll 1$, and
this procedure then results in the system of coupled linear equations
\begin{equation}
\left(g  + \widetilde{G}_{1} \right) \alpha_{1} -  h \alpha_{2} =
\frac{\beta e^{2} J}{{\cal G}} P_{1} \; ,
\label{eq:69}
\end{equation}
\begin{equation}
h \alpha_{1} - \left( g + \widetilde{G}_{2} \right) \alpha_{2} = \frac{\beta
e^{2} J}{{\cal G}} P_{2} \; ,
\label{eq:69abc}
\end{equation}
where
\begin{equation}
\widetilde{G}_{1,2} = \frac{G_{1,2}}{{\cal G}} \; .
\label{eq:90klm}
\end{equation}
The solutions of Eqs.\ (\ref{eq:69}) and (\ref{eq:69abc}) are found to be
\begin{equation}
\alpha_{1} = \frac{\beta e^{2} J}{{\cal G} \Delta} \left[ \left( g +
\widetilde{G}_{2} \right) P_{1} - h P_{2} \right] \; ,
\label{eq:72abc}
\end{equation}
\begin{equation}
\alpha_{2}  = \frac{\beta e^{2} J}{{\cal G} \Delta} \left[ h P_{1} - \left(
g + \widetilde{G}_{1} \right) P_{2} \right] \; ,
\label{eq:72}
\end{equation}
where
\begin{equation}
\Delta = \left( g + \widetilde{G}_{1} \right) \left( g +  \widetilde{G}_{2}
\right) - h^{2}  \; .
\label{eq:72ijk}
\end{equation}
Expressions (\ref{eq:72abc}) and (\ref{eq:72}) determine the spin fraction
excesses $\alpha_{1}$ and $\alpha_{2}$ in terms of the current $J$, of the
polarizations $P_{1}$ and $P_{2}$ in the left and right ferromagnet,
respectively, and of material parameters, such as the conductivities
$\sigma_{1,2}$ and the spin diffusion lengths $L_{s}^{(1,2)}$ of the
ferromagnets (via $\widetilde{G}_{1,2}$), and the momentum relaxation length
$l$ and the spin relaxation length $l_{s}$ of the semiconductor as well as its
length $S$ (via $g$ and $h$) and the equilibrium density $\bar{n}$ (via ${\cal
G}$).  Since the quantities $\alpha_{1,2}$ are proportional to the current $J$,
the current spin polarization $P_{J}(x)$ is independent of $J$, while the
density spin polarization $P_{n}(x)$ is proportional to $J$.

The current spin polarization along the entire heterostructure, $P_{J}(x)$, is
now obtained as follows.  In the semiconductor, it is given by expression
(\ref{eq:40c}), with $C_{1,2}$ calculated from $\alpha_{1,2}$ as explained
there.  In the ferromagnets, the expressions for the current spin polarization
are provided by Eqs.\ (\ref{eq:57}) and (\ref{eq:57zzz}), respectively, where
the quantities $\mu_{-}(x_{1}^{-})$ and $\mu_{-}(x_{2}^{+})$ are calculated
from Eq.~(\ref{eq:karol}) and from its analogue for $\mu_{-}(x_{2}^{+})$,
respectively.  Analogously, the density spin polarization $P_{n}(x)$ in the
semiconductor is given by expression (\ref{eq:40rice}).  We do not write down
the density spin polarizations in the ferromagnets, but only mention that they
do not, in general, match the polarizations $P_{n}(x_{1}^{+})$ and
$P_{n}(x_{2}^{-})$ on the semiconductor sides of the interfaces.

\begin{figure}
\vspace*{0.0cm}
\epsfysize=8.0cm
\epsfbox[50 143 445 553]{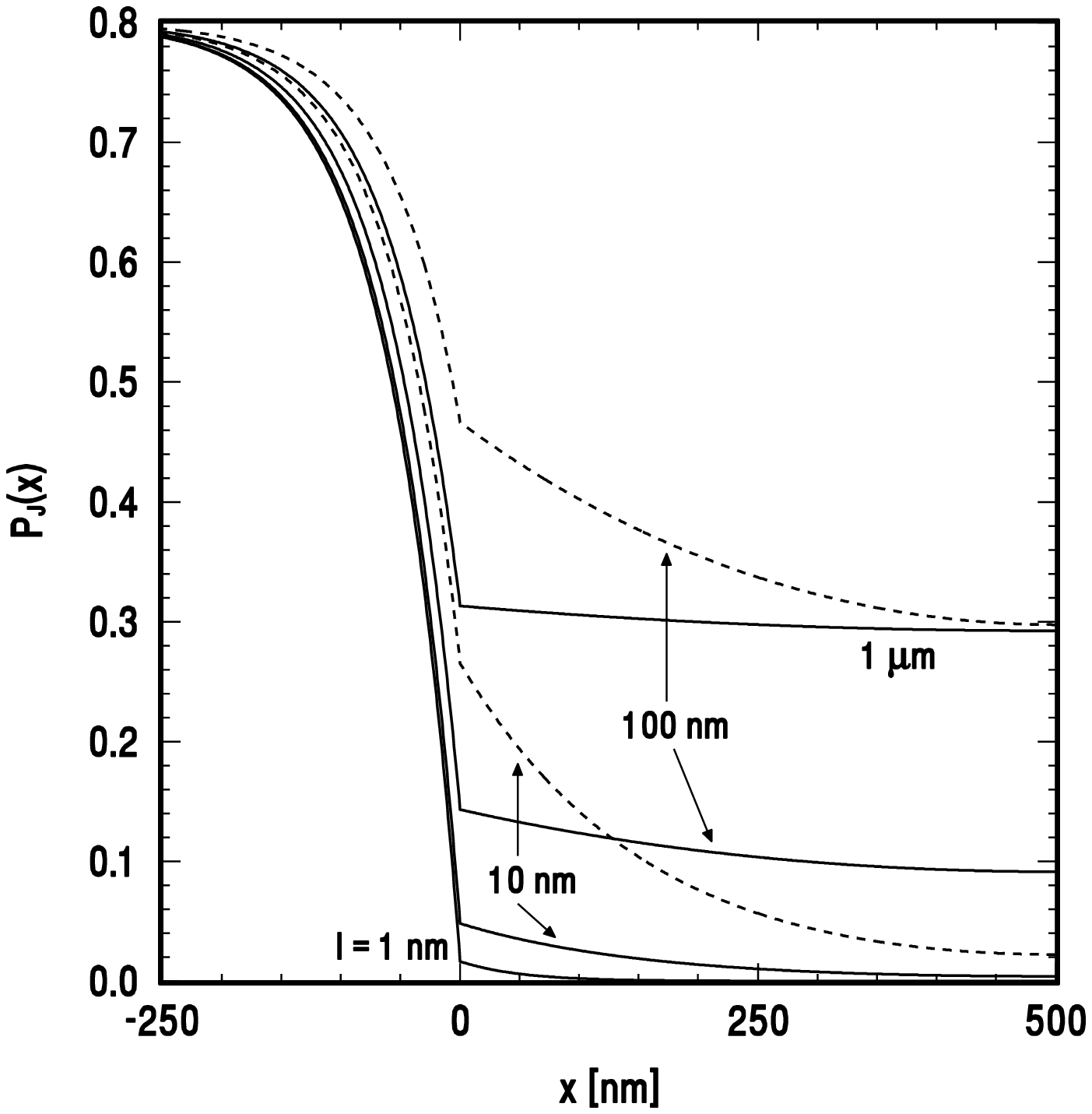}
\vspace*{0.0cm}
\caption{The zero-bias ($\epsilon \rightarrow 0$) current spin polarization
$P_{J}(x)$ along a symmetric ferromagnet/semi\-conduc\-tor/ferro\-mag\-net
heterostructure with $S = 1$ $\mu$m for the indicated values of the momentum
relaxation length $l$, calculated from Eqs.\ (\ref{eq:57}), (\ref{eq:57zzz}),
and (\ref{eq:40c}) with $x_{1} = 0$ and $x_{2} = S$.  The solid curves
correspond to zero interface resistance, the dashed curves to interface
resistances of $10^{-7}$ $\Omega$ cm$^{2}$ for spin-up electrons and $5 \times
10^{-7}$ $\Omega$ cm$^{2}$ for spin-down electrons, respectively.  For the
remaining parameter values, see text. \vspace*{-0.3cm}
}
\label{fig:5}
\end{figure}

In order to demonstrate the effect of the transport mechanism (characterized by
the magnitude of the ratios $l/l_{s}$ and $l/S$), we show in Fig.\ \ref{fig:5}
the zero-bias current spin polarization $P_{J}(x)$ for a symmetric
ferromagnet/semiconductor/ferromagnet heterostructure with sample length $S =
1$ $\mu$m at $T = 300$ K as a function of $x$ for various values of the
momentum relaxation length $l$.  For the parameters of the ferromagnets, we
adopt from Ref.\ \onlinecite{yuf02a} the values $\sigma_{1} = \sigma_{2} =
10^{3}$ $\Omega^{-1}$ cm$^{-1}$ for the bulk conductivities and $L_{s}^{(1)} =
L_{s}^{(2)} = 60$ nm for the spin diffusion lengths; the bulk polarizations are
chosen as $P_{1} = P_{2} = 0.8$.  For the material parameters of the
semiconductor, we take the values $m^{\ast} = 0.067 m_{e}$ for the effective
electron mass, $l_{s} = 2.5$ $\mu$m for the ballistic spin relaxation length
(corresponding to n-doped GaAs; see Refs.~\onlinecite{kim01} and
\onlinecite{bec05}), and $\bar{n} = 5.0 \times 10^{17}$ cm$^{-3}$ for the
equilibrium electron density.  Clearly, in a specific semiconducting system,
the value of the momentum relaxation length $l$ is fixed.  Therefore, when
varying $l$, we are considering the above parameter values to be representative
for a whole class of semiconductors (regarded as nondegenerate; at room
temperature, this should be an acceptable working hypothesis\cite{ami04}) that
differ in the strength of impurity and phonon scattering and hence in the
magnitude of $l$.

The momentum relaxation length $l$ affects the results shown in Fig.\
\ref{fig:5} in a twofold way.  (i) It determines the conduction in the
semiconductor.  For small values of $l$, the conductance of the latter is
small, and thus the conductance mismatch with the ferromagnets is large,
leading to a small injected current spin polarization $P_{J}(0)$.  (ii) It
determines the generalized spin diffusion length $L =
[ll_{s}/(1+l/l_{s})]^{1/2}$, which acts as the polarization decay length, so
that for small $l$ the polarization dies out rapidly inside the semiconductor.
The degree of polarization may be raised considerably all along the
semiconductor when the value of $l$ is increased up to a length of the order of
the sample length, in which case the ballistic component becomes prevalent.
Figure \ref{fig:5} also shows that, by introducing appropriately chosen
spin-selective interface resistances, one may offset the suppression of the
injected polarization due to the conductance mismatch for small $l$; however,
the rapid decay of the polarization inside the semiconductor cannot be
prevented in this way.

\begin{figure}[t]
\vspace*{0.0cm}
\epsfysize=8.0cm
\epsfbox[50 143 445 553]{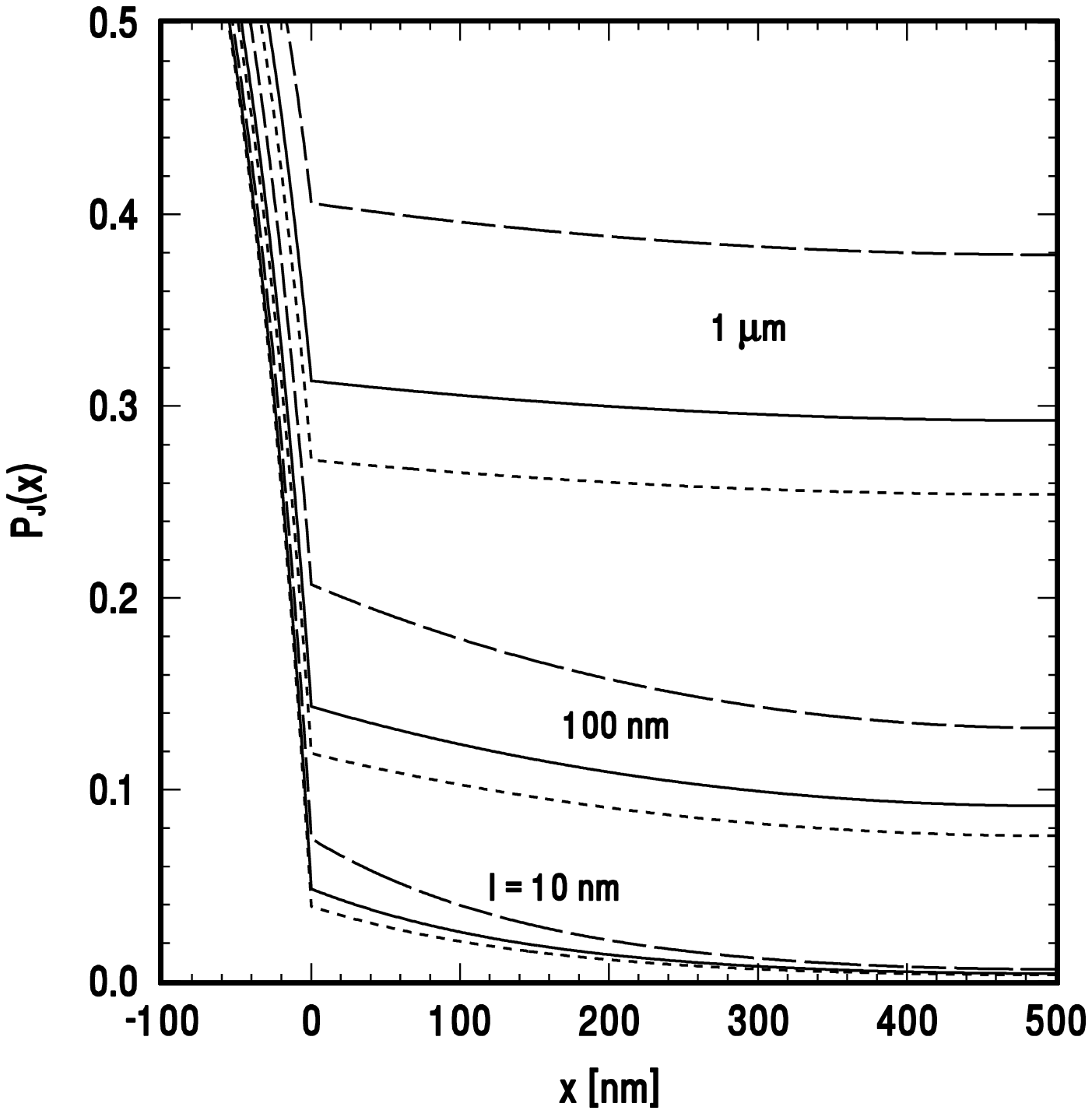}
\vspace*{0.0cm}
\caption{The zero-bias current spin polarization $P_{J}(x)$ for the case of
Fig.\ \protect{\ref{fig:5}}, for zero interface resistance and  various values
of the momentum relaxation length $l$ and of the equilibrium density $\bar{n}$.
Short-dashed curves:\ $ \bar{n} = 4 \times 10^{17}$ cm$^{-3}$; full curves:\
$\bar{n} =  5 \times 10^{17}$ cm$^{-3}$; long-dashed curves:\  $\bar{n} =
8 \times 10^{17}$ cm$^{-3}$.\vspace*{-0.3cm}
}
\label{fig:6}
\end{figure}

\begin{figure}[t]
\vspace*{0.0cm}
\epsfysize=8.0cm
\epsfbox[50 143 445 553]{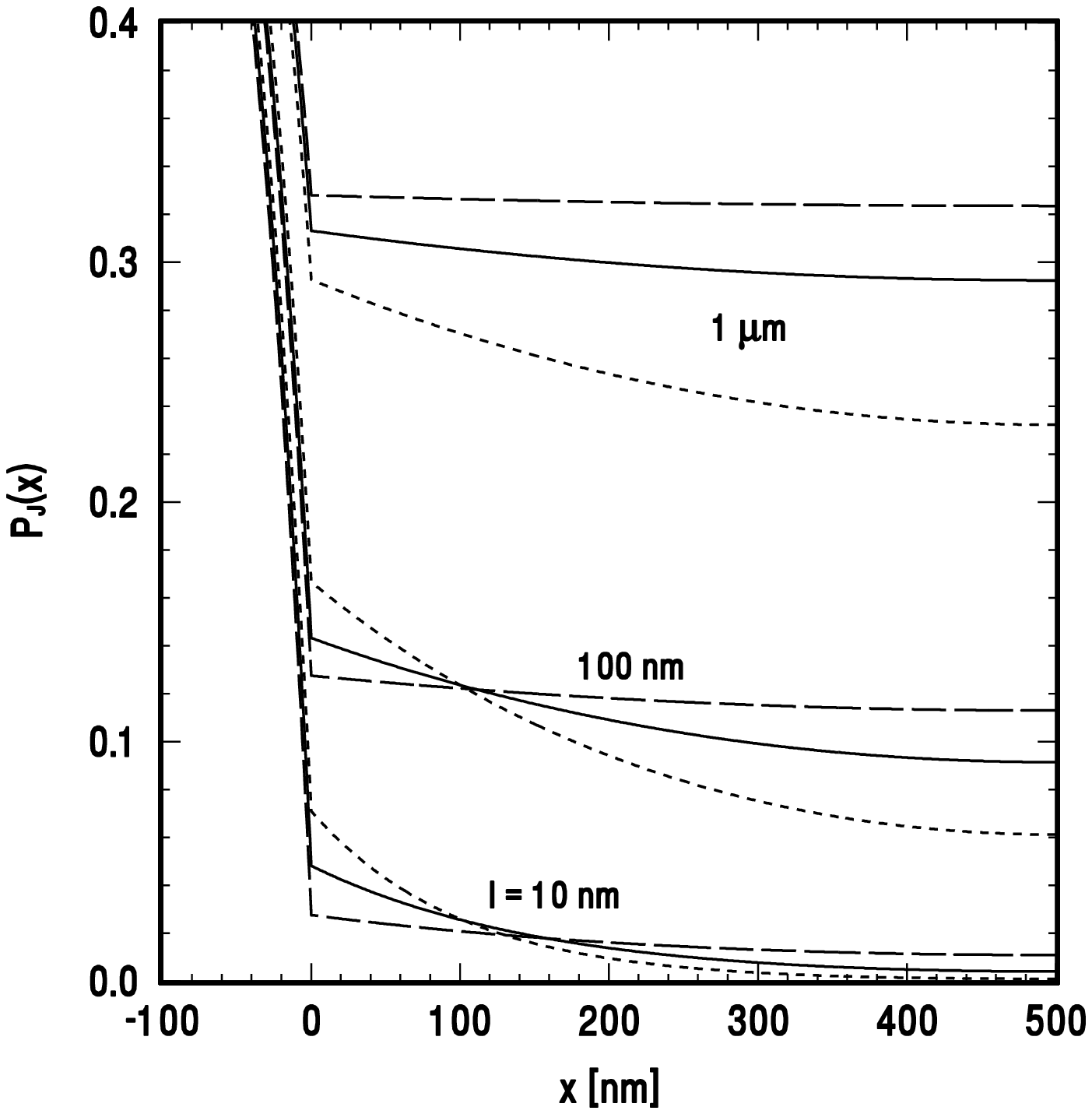}
\vspace*{0.0cm}
\caption{The zero-bias current spin polarization $P_{J}(x)$ for the case of
Fig.\ \protect{\ref{fig:5}}, for zero interface resistance and  various values
of the momentum relaxation length $l$ and of the spin relaxation length
$l_{s}$. Short-dashed curves:\ $l_{s} =  1$ $\mu$m; full curves:\
$l_{s} =  2.5$ $\mu$m; long-dashed curves:\  $l_{s} = 10$ $\mu$m.
\vspace*{-0.3cm}
}
\label{fig:7}
\end{figure}

For the case of Fig.\ \ref{fig:5}, we show in Figs.\ \ref{fig:6} and
\ref{fig:7}, respectively, the zero-bias current spin polarization $P_{J}(x)$
for various values of the equilibrium density $\bar{n}$ and the spin relaxation
length $l_{s}$.  It is seen that varying $\bar{n}$ has about the same
overall effect on $P_{J}(x)$ as varying the momentum relaxation length $l$,
whereas varying $l_{s}$ affects mainly the rate of decay of $P_{J}(x)$.

\subsection{Injected spin polarization for field-driven transport}

We introduce the ``injected spin polarization" as the spin polarization at
one of the interfaces, e.g., at $x=x_{1}$, generated by the bulk polarization
$P_{1}$ of the left ferromagnet regardless of the influence of the right
ferromagnet.  More precisely, we define the injected current spin polarization
as the current spin polarization $P_{J}(x_{1}^{+})$ given by Eq.\
(\ref{eq:40a}) in the limit $S/L \rightarrow \infty$.  Similarly, the injected
density spin polarization is defined as the polarization $P_{n}(x_{1}^{+})$ of
Eq.\ (\ref{eq:40b}) in the same limit.  The injected spin polarization at $x =
x_{1}^{+}$ provides the initial value of the left-generated polarization in the
semiconductor, which propagates into the region $x > x_{1}$ while being
degraded by the effect of spin relaxation.

We now consider the injected spin polarization for electron transport driven by
an external electric field, i.e., a potential profile of the form
(\ref{eq:38aaaa}).

\subsubsection{General case}

In order to obtain the spin transport function $A(x)$, we have to solve Eq.\
(\ref{eq:38csu}) numerically under the condition $A(x) \propto
\exp(-x/\lambda)$ for $x \rightarrow \infty$.  The decay length $\lambda$ is
determined by solving Eq.\ (\ref{eq:38csu}) in the range $x - x_{1} \gg
(\epsilon + 1/l)^{-1}$ where the function $b(x)$ in the coefficient functions
$b_{0}(x)$, $b_{1}(x)$, and $b_{2}(x)$ can be disregarded,
\begin{eqnarray}
\lambda &=& \left\{ \frac{\epsilon}{2} + \left[ \frac{\epsilon^{2}}{4} +
\frac{1+ \epsilon \tilde{l}}{\tilde{l}^{2}} - \frac{1 + \epsilon l}{l
\bar{l}} \frac{2 + \epsilon \bar{l}}{2 + \epsilon l} \right]
^{1/2} \right\} ^{-1} \; . \nonumber \\
\label{eq:80}
\end{eqnarray}
It can be shown that for any combination of parameter values, $\lambda$ is
a real number.

For calculating the injected current spin polarization from
Eq.~(\ref{eq:40a}), we determine the thermoballistic spin-polarized current at
the interface, $J_{-}(x_{1}^{+})$, from Eq.~(\ref{eq:23}).  Using
Eq.~(\ref{eq:21}) and fixing the normalization of the function $A(x)$ in terms
of $A_{1}$ with the help of Eq.~(\ref{eq:38b}), we find
\begin{equation}
J_{-}(x_{1}^{+})  = v_{e} \bar{n} \, \Gamma_{J} \, \alpha_{1}
\; , \label{eq:81}
\end{equation}
where
\begin{equation}
\Gamma_{J}  = \frac{A(x_{1}^{+}) - \bar{A}}{A(x_{1}^{+}) -
\bar{A}/2}
\label{eq:82}
\end{equation}
and
\begin{equation}
\bar{A} =  \int_{x_{1}}^{\infty} \frac{dx}{l} \, e^{-(x-x_{1})/\tilde{l}} \,
A(x) \; . \label{eq:83}
\end{equation}
To find the total thermoballistic current at the interface, $J(x_{1}^{+})$, we
go back to Eq.\ (\ref{eq:5e}).  Expressing the current $J$ in the form
\begin{equation}
J =  \frac{1}{\tilde{\chi}} \, v_{e} \bar{n} \; ,
\label{eq:9zyx}
\end{equation}
which follows, for $\epsilon > 0$ and $S/L \rightarrow \infty$, from the
current-voltage characteristic (\ref{eq:9xy}) with $\beta e V = \epsilon S$ and
$N_{c} \, \exp(-\beta E_{p}) = \bar{n}$, we obtain
\begin{equation}
J(x_{1}^{+}) = \frac{\kappa}{\tilde{\chi}} \, v_{e} \bar{n}  \; .
\label{eq:9zyan}
\end{equation}
This expression is conveniently evaluated by using for $\kappa$ and
$\tilde{\chi}$ the closed-form representations
\begin{equation}
\kappa = \frac{1 + \epsilon l}{2 + \epsilon l} \; , \; \; \tilde{\chi} =
\frac{(1 + \epsilon l)^{2}}{\epsilon l (2+ \epsilon l)} \; ,
\label{eq:69uvw}
\end{equation}
which have been inferred from the results of systematic numerical calculations
for fixed $\epsilon l > 0$ and very large values of $S/L$. For the injected
current spin polarization, we now find
\begin{equation}
P_{J}(x_{1}^{+}) = \frac{J_{-}(x_{1}^{+})}{J(x_{1}^{+})}  =
\frac{ \tilde{\chi}}{\kappa} \, \Gamma_{J} \, \alpha_{1}  \; ,
\label{eq:81ijk}
\end{equation}
which, by continuity, is equal to $P_{J}(x_{1})$.  Setting the right-hand side
of Eq.\ (\ref{eq:81ijk}) equal to expression (\ref{eq:57angi}) [or to the
more general expression including spin-selective interface resistances; see
the remark following Eqs.~(\ref{eq:61a}) and (\ref{eq:61abc})] for
the injected spin polarization in terms of the contact parameters, we arrive at
\begin{equation}
P_{1}  - \frac{\widetilde{G}_{1}}{2}  \tilde{\chi} \ln \left( \frac{1+
\alpha_{1}}{1 -\alpha_{1}} \right) = \frac{ \tilde{\chi}}{\kappa} \,
\Gamma_{J} \, \alpha_{1}  \; .
\label{eq:69rst}
\end{equation}
This is a nonlinear equation for $\alpha_{1}$ which is to be solved for given
values of the parameters $\epsilon$, $P_{1}$, $G_{1}$, $\bar{n}$, $l$, and
$l_{s}$.

\begin{figure}[t]
\vspace*{0.0cm}
\epsfysize=8.0cm
\epsfbox[50 143 445 553]{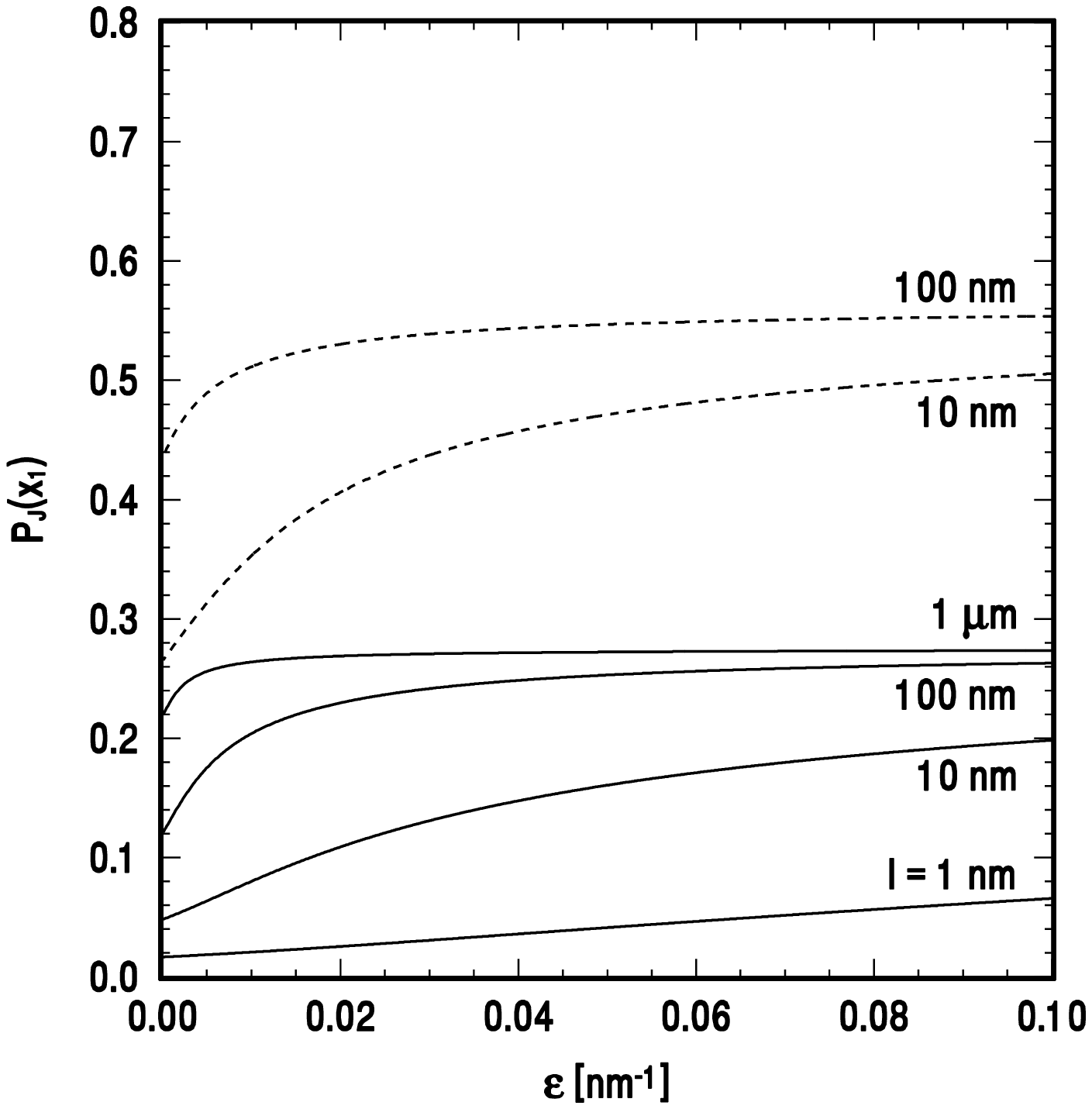}
\vspace*{0.0cm}
\caption{The injected current spin polarization $P_{J}(x_{1})$ for $S/L
\rightarrow \infty $ as a function of the electric-field parameter $\epsilon$
for the indicated values of the momentum relaxation length $l$.  The solid
curves correspond to zero interface resistance, the dashed curves to interface
resistances of $10^{-7}$ $\Omega$ cm$^{2}$ for spin-up electrons and $5 \times
10^{-7}$ $\Omega$ cm$^{2}$ for spin-down electrons, respectively.  For the
remaining parameter values, see text. \vspace*{-0.3cm}
}
\label{fig:8}
\end{figure}

Turning to the calculation of the injected density spin polarization, we
determine the thermoballistic spin-polarized density at the interface,
$n_{-}(x_{1}^{+})$, from Eq.~(\ref{eq:24}), using again Eqs.\
(\ref{eq:21}) and (\ref{eq:38b}), and obtain
\begin{equation}
n_{-}(x_{1}^{+}) = \frac{\bar{n}}{2} \, \Gamma_{n} \,
\alpha_{1} \; ,
\label{eq:82pqqq}
\end{equation}
where
\begin{equation}
\Gamma_{n}  = \frac{A(x_{1}^{+})}{A(x_{1}^{+}) - \bar{A}/2} \; .
\label{eq:82pqr}
\end{equation}
For the total thermoballistic density at the interface, $n(x_{1}^{+})$, we find
from Eq.\ (\ref{eq:11ahab}), using Eqs.\ (\ref{eq:12stu}), (\ref{eq:11aha}),
and (\ref{eq:9zyx}),
\begin{equation}
n(x_{1}^{+})  = \bar{n} \left(1 - \frac{\kappa J}{2 v_{e} \bar{n}}
\right) =\bar{n} \left(1 - \frac{\kappa }{2 \tilde{\chi}}
\right) \; .
\label{eq:82pqq}
\end{equation}
The injected density spin polarization now follows as
\begin{equation}
P_{n}(x_{1}^{+})  = \frac{n_{-}(x_{1}^{+})}{n(x_{1}^{+})} =
\frac{\tilde{\chi}}{2 \tilde{\chi} - \kappa} \, \Gamma_{n} \, \alpha_{1} \; ,
\label{eq:81zyx}
\end{equation}
where the spin fraction excess $\alpha_{1}$ is again to be determined by
solving Eq.\ (\ref{eq:69rst}).

Figure \ref{fig:8} shows the injected current spin polarization $P_{J}(x_{1})$
for $S/L \rightarrow \infty$ as a function of the electric-field parameter
$\epsilon$ for various values of the momentum relaxation length $l$; the
remaining parameter values are the same as in Fig.\ \ref{fig:5}.  In
calculating $P_{J}(x_{1})$ from Eq.\ (\ref{eq:81ijk}), we have used $\kappa =
1$.  This choice has been made because expression (\ref{eq:81ijk}) with
$\kappa$ given by Eq.\ (\ref{eq:69uvw}) does not represent a meaningful
injected polarization in the diffusive limit (see below); instead, one must set
$\kappa = 1$ in this limit.  For simplicity, we have used this value
throughout.  In conformity with the drift-diffusion results of Ref.\
\onlinecite{yuf02a}, the injected polarization generally rises with increasing
$\epsilon$; however, as in Fig.\ \ref{fig:5}, the main effect is due to the
variation of $l$.

\subsubsection{Diffusive regime}

In order to relate our treatment of the injected spin polarization at
ferromagnet/semi\-conductor interfaces to previous treatments within the
drift-diffusion model, in particular to that of Yu and Flatt\'{e},\cite{yuf02a}
we consider the diffusive regime, $l/l_{s} \ll 1$ and $\epsilon l \ll 1$, in
some detail. In that regime, the spin transport function $A(x)$ is determined
by Eq.\ (\ref{eq:38d}), whose solution is $A(x) \propto \exp(-x/
L_{s}^{\epsilon})$, with the field-dependent spin diffusion length
$L_{s}^{\epsilon}$ given by
\begin{equation}
\frac{1}{L_{s}^{\epsilon}} =  \frac{\epsilon}{2} + \left(
\frac{\epsilon^{2}}{4} + \frac{1}{L_{s}^{2}} \right) ^{1/2} \; .
\label{eq:80hui}
\end{equation}
We then obtain $\Gamma_{J} = 2 l/L_{s}^{\epsilon}$ and $\Gamma_{n} = 2$.
Furthermore, from Eqs.\ (\ref{eq:69uvw}) for $\epsilon l \ll 1$, we find
$\kappa = 1/2$ and $\tilde{\chi} = 1/2\epsilon l$.

At this point, some analysis is required regarding the definition of the
injected current spin polarization in the diffusive regime. In the definition
introduced above, first the functions $J_{-}(x)$ and $J(x)$ are evaluated for
$x \rightarrow x_{1}^{+}$, and subsequently the diffusive limit is
approached.  This procedure results, in particular, in the value $\kappa
= J(x_{1}^{+})/J = 1/2$.  A closer look at the function $J(x)/J$ (see Fig.\
\ref{fig:3}), however, shows that in the diffusive regime this function is
virtually equal to unity inside the semiconducting sample and tends to smaller
values only within a (very short) distance of order $l$ from the interfaces.
Therefore, it is indicated here to define the injected current spin
polarization in terms of a position $x > l$ inside the sample, where $J(x)/J =
1$ is the relevant value for the propagation of the spin polarization into the
semiconductor.  Thus, in the diffusive regime, we adopt the effective value
$\kappa = 1$ in the calculation of the injected spin polarization.  While in
the ballistic limit (and now also in the diffusive regime) the choice $\kappa
= 1$ is unique, in the range of intermediate $l$-values a meaningful definition
of the injected spin polarization requires an appropriate choice of the
position inside the sample at which the thermoballistic current and
spin-polarized current are to be evaluated.

With the choice $\kappa = 1$, the injected current spin polarization in the
diffusive regime is obtained from  Eq.\ (\ref{eq:81ijk}) as
\begin{equation}
P_{J}(x_{1})  = \frac{1}{\epsilon L_{s}^{\epsilon}} \, \alpha_{1} \; ,
\label{eq:81ghj}
\end{equation}
where the spin fraction excess $\alpha_{1}$ is now to be calculated from Eq.\
(\ref{eq:69rst}) with $\kappa = 1$.  Since $\kappa \ll \tilde{\chi}$ for
$\epsilon l \ll 1$, the injected density spin polarization in the diffusive
regime follows from Eq.\ (\ref{eq:81zyx}) as
\begin{equation} P_{n}(x_{1}^{+})  = \alpha_{1} \; ,
\label{eq:81jhg}
\end{equation}
where $\alpha_{1}$ again must  be calculated from Eq.\ (\ref{eq:69rst}) with
$\kappa = 1$.

Comparing our results for the injected spin polarization in the diffusive
regime to the results of Yu and Flatt\'{e}\cite{yuf02a} based on standard
drift-diffusion theory, we find that the field-dependent spin diffusion length
$L_{s}^{\epsilon}$ given by Eq.\ (\ref{eq:80hui}) agrees with the ``up-stream'' spin
diffusion length $L_{u}$ given by Eq.\ (2.23b) of Ref.~\onlinecite{yuf02a},
provided the intrinsic spin diffusion length $L$ of that reference is
identified with the spin diffusion length $L_{s} = \sqrt{l l_{s}}$ of the
present work.  Then, by expressing the conductivity of the semiconductor in
Eq.\ (3.5) of Ref.~\onlinecite{yuf02a} (with the interface resistances set
equal to zero) as $\sigma_{s} = 2 \beta e^{2} v_{e} \bar{n}_{1} l$, we
recognize the equivalence of that equation with our Eq.\ (\ref{eq:69rst}) in
the diffusive regime.  This, in turn, implies that the injected current and
density spin polarizations of either work are formally identical.  Numerical
calculations have confirmed this result.

\subsubsection{Zero-bias limit}

We now consider the injected current spin polarization in the zero-bias limit,
in which $|\alpha_{1}| \ll 1$ and $J(x) = J$, i.e., $\kappa = 1$.  Here, the
spin transport function $A(x)$ is determined by Eq.\ (\ref{eq:30}), i.e., $A(x)
\propto \exp(-x/L)$, so that $\Gamma_{J} = 2 \gamma/(1 + \gamma) \equiv
\gamma_{J}$ and $\Gamma_{n} = 2/(1 + \gamma) \equiv \gamma_{n}$.  From Eq.\
(\ref{eq:69rst}), we then have
\begin{equation}
\alpha_{1} =  \frac{P_{1}}{ \tilde{\chi} \gamma_{J}  (1 + G_{1}/\gamma_{J}
{\cal G})}  \; .
\label{eq:69ahoi}
\end{equation}
Combining this with Eq.\ (\ref{eq:81ijk}), we find for the injected current
spin polarization
\begin{equation}
P_{J}(x_{1}) = \frac{1}{1 + G_{1}/\gamma_{J} {\cal G}} \, P_{1} \; .
\label{eq:70xy}
\end{equation}
It is instructive to consider expression (\ref{eq:70xy}) in the diffusive and
ballistic regimes.

In the {\em diffusive} regime $l/l_{s} \ll 1$, we have $\gamma_{J} = 2
\sqrt{l/l_{s}}$ and, therefore,
\begin{equation}
P_{J}(x_{1}) =  \frac{1}{1 + G_{1}/{\cal G}_{0}} \, P_{1} \; .
\label{eq:75}
\end{equation}
Here, ${\cal G}_{0} = 2 {\cal G} \sqrt{l/l_{s}} = \sigma_{0}/L_{s}$, where Eq.\
(\ref{eq:5aaa}) has been used, and $\sigma_{0} = 2 {\cal G} l$ is the
conductivity of the semiconductor [see the remarks following Eq.\
(\ref{eq:11heid})].  The quantity ${\cal G}_{0}$ is seen to be the
semiconductor analogue of the ferromagnet parameter $G_{1}$ defined by Eq.\
(\ref{eq:57xyz}). Choosing $P_{1} = 0.8$ and adopting the values\cite{yuf02a}
$\sigma_{1} = 10^{3}$ $\Omega^{-1} {\rm cm}^{-1}$, $L_{s}^{(1)} = 60$ nm,
$\sigma_{0} = 10$ $\Omega^{-1} {\rm cm}^{-1}$, and $L_{s} = 2$ $\mu$m, we have
$G_{1}/{\cal G}_{0} = 1.2 \times 10^{3}$ and hence $P_{J}(x_{1}) \approx 0.6
\times 10^{-3}$. The large value of the ratio $G_{1}/{\cal G}_{0}$
reflects the ``conductance mismatch'' which appears to be the determining
parameter of the injected spin polarization in the diffusive
regime.\cite{sch02,sch05,sch00}

On the other hand, in the {\em ballistic} limit $l/l_{s} \rightarrow \infty$,
we have $\gamma_{J} = \gamma = 1$,
so that
\begin{equation}
P_{J}(x_{1})  = \frac{1}{1 + G_{1}/{\cal G}} \, P_{1} \; .
\label{eq:79y}
\end{equation}
Here, the Sharvin interface conductance ${\cal G}$ takes the place of the
quantity ${\cal G}_{0}$ in Eq.\ (\ref{eq:75}).  Assuming $m^{*} = 0.067 m_{e}$
and $T = 300$ K, we have ${\cal G} = 3.2 \times 10^{11}$ $\Omega^{-1}$m$^{-2}$
for $\bar{n} = 5 \times 10^{17}$ cm$^{-3}$ and ${\cal G} = 0.64 \times 10^{10}$
$\Omega^{-1}$m$^{-2}$ for $\bar{n} = 10^{16}$ cm$^{-3}$.  This results in
$P_{J}(x_{1}) \approx 0.3$ and $0.8 \times 10^{-2}$, respectively.  The first
example, where the large Sharvin interface conductance entails a large injected
spin polarization, is fictitious since the high doping concentration needed to
obtain an electron density of $5 \times 10^{17}$ cm$^{-3}$ (for example, in
GaAs) would imply such small values of the momentum relaxation length $l$ that
ballistic transport is all but ruled out.  Only semiconducting materials with
unusually large mobilities would make this a realistic case.  The second
example with the lower electron density of $10^{16}$ cm$^{-3}$ would be more
favorable to a ballistic transport mechanism, but leads to a very small
injected spin polarization; this confirms the conclusion of Kravchenko and
Rashba\cite{kra03} stating that spin injection is suppressed even in the
ballistic regime unless spin-selective interface resistances are introduced.

\section{Concluding remarks}

We have developed a unified semiclassical theory of spin-polarized electron
transport in heterostructures formed of a nondegenerate semiconductor and two
ferromagnetic contacts.  In this theory, the spin polarization inside the
semiconductor is obtained for a general transport mechanism that covers the
whole range between the purely diffusive and purely ballistic mechanisms and is
controlled by the momentum relaxation length of the electrons.

The basis of the present work is provided by our previously developed unified
model of (spinless) electron transport in semiconductors, in which diffusive
and ballistic transport are combined in the concept of the thermoballistic
electron current.  As a prerequisite to the extension of the spinless unified
model to spin-polarized transport, we have modified and completed its
formulation in such a way that an unambiguous description of electron transport
in terms of a uniquely defined thermoballistic chemical potential is achieved.
From the chemical potential, the unique thermoballistic current and density are
obtained; numerical calculations show that, for typical parameter values, the
thermoballistic current is close to the physical current.

In order to treat spin-polarized transport in semiconductors within the unified
description, we have introduced a thermoballistic spin-polarized current and a
thermoballistic spin-polarized density by allowing spin relaxation to take
place during the ballistic electron motion.  These are expressed in terms of a
spin transport function which comprises in a compact form the information
contained in the spin-resolved thermoballistic chemical potentials.  Using the
balance equation that connects the thermoballistic spin-polarized current and
density, we have derived an integral equation for the spin transport function,
from which the latter can be calculated in terms of its values at the
interfaces of the semiconductor with the contacts.  The spin transport function
determines, in conjunction with the spin-independent thermoballistic chemical
potential, all spin-dependent quantities in the semiconductor, in particular,
the position dependence of the current and density spin polarization.  The spin
polarization all across a ferromagnet/semiconductor heterostructure is
determined by making use of the continuity of the current spin polarization at
the contact-semiconductor interfaces and connecting the spin-resolved chemical
potentials there.  Thereby, a unified description of spin-polarized transport
emerges that provides a basis for the systematic study of the interplay of spin
relaxation and transport mechanism in heterostructures relevant to spintronic
applications.

To interpret the formalism developed here and to relate it to previous, less
general formulations, we have considered spin-polarized electron transport in a
homogeneous semiconductor without space charge, driven by an external electric
field.  Within a judicious approximation, the integral equation for the spin
transport function can then be reduced to a second-order differential equation
which generalizes the standard spin drift-diffusion equation to the case of
arbitrary values of the ratio of momentum to spin relaxation length.  In the
zero-bias limit, the position dependence of the spin polarizations across a
heterostructure is obtained in closed form.

The generalized spin drift-diffusion equation has been used in calculations of
the current spin polarization across a symmetric
ferromagnet/semiconductor/ferromagnet heterostructure with material parameters
in the range of interest for spintronic devices.  The dependence on the
transport mechanism in the semiconductor has been exhibited by varying the
momentum relaxation length over several orders of magnitude.  It was found that
the ballistic regime favors sizeable (large) spin polarizations.  The same
picture emerges from calculations of the injected current spin polarization as
a function of an applied electric field.  While the field also serves to raise
the polarization in the semiconductor, the main effect still is due to the
variation of the momentum relaxation length, i.e., to the influence of the
ballistic component of the transport mechanism.  In order to exploit the
potentiality of varying the transport mechanism with the aim to improve the
efficiency of spintronic devices, the identification and design of novel
semiconducting materials is called for.

In the present work, emphasis has been placed on a careful elaboration of the
formalism underlying the unified description of spin-polarized electron
transport in ferromagnet/semiconductor heterostructures.  In the illustrative
calculations, we have restricted ourselves to the simplest cases.  In future
work, applications of the present formalism will have to be based on the
solution of the fundamental integral equation for the spin transport function
in its general form.  These should include the treatment of magnetic
semiconducting samples (characterized, in the unified description, by nonzero
values of the relaxed spin fraction excess) and of interface barriers, like
Schottky or tunnel barriers (represented by appropriately chosen potential
profiles).  As to possible extensions of the theory, setting up a formalism for
the treatment of degenerate semiconductors appears to have first priority.

\appendix*

\section{Unique thermoballistic functions}

In this Appendix, we present details of the construction of a unique
thermoballistic chemical potential $\mu(x)$, current $J(x)$, and density $n(x)$
in terms of the solutions $\chi_{1}(x)$ and $\chi_{2}(x)$ of Eqs.\
(\ref{eq:11bb}) and (\ref{eq:11bbb}), respectively.

Evaluating expression (\ref{eq:11a}) with the function ${\cal J}_{1}(x)$
following from the solution of Eq.\ (\ref{eq:11bb}) [case (i)], with ${\cal
J}_{2}$ set equal to ${\cal J}_{1}(x_{2})$, we obtain the thermoballistic
current
\begin{eqnarray}
&\ & \hspace*{-0.9cm} J_{1}(x) = w_{1}(x_{1},x_{2};l) \; [{\cal
J}_{1} - {\cal J}_{1}(x_{2})] \nonumber \\ &+& \int_{x_{1}}^{x}\frac{dx'}{l} \,
w_{2}(x',x_{2};l) \, [{\cal J}_{1}(x') - {\cal J}_{1}(x_{2})] \nonumber \\  &+&
\int_{x}^{x_{2}} \frac{dx'}{l} \, w_{2}(x_{1},x';l) \, [{\cal J}_{1} - {\cal
J}_{1}(x')] \nonumber \\ &+& \int_{x_{1}}^{x} \frac{dx'}{l} \int_{x}^{x_2}
\frac{dx''}{l} \, w_{3}(x',x'';l) \, [{\cal J}_{1} (x') - {\cal J}_{1}(x'')]
\nonumber \\
\label{eq:11ax}
\end{eqnarray}
and, similarly, for case (ii), the current
\begin{eqnarray}
&\ & \hspace*{-0.9cm} J_{2}(x) = w_{1}(x_{1},x_{2};l) \; [{\cal
J}_{2}(x_{1}) - {\cal J}_{2}] \nonumber \\ &+& \int_{x_{1}}^{x}\frac{dx'}{l} \,
w_{2}(x',x_{2};l) \, [{\cal J}_{2}(x') - {\cal J}_{2}] \nonumber \\ &+&
\int_{x}^{x_{2}} \frac{dx'}{l} \, w_{2}(x_{1},x';l) \, [{\cal J}_{2}(x_{1}) -
{\cal J}_{2}(x')] \nonumber \\ &+& \int_{x_{1}}^{x} \frac{dx'}{l}
\int_{x}^{x_2} \frac{dx''}{l} \, w_{3}(x',x'';l) \, [{\cal J}_{2}(x') - {\cal
J}_{2}(x'')] \; , \nonumber \\
\label{eq:11ay}
\end{eqnarray}
which are not, in general, equal.  This ambiguity is removed by introducing a
unique thermoballistic current $J^{(u)}(x)$ as a superposition of the
currents $J_{1}(x)$ and $J_{2}(x)$,
\begin{equation}
J^{(u)}(x) = \hat{a}_{1} \, \frac{ J}{J_{1}} \, J_{1}(x) + \hat{a}_{2}\, \frac{
J}{J_{2}} \, J_{2}(x) \; ,
\label{eq:11}
\end{equation}
where
\begin{equation}
J_{1,2} = \frac{1}{x_{2} - x_{1}} \int_{x_{1}}^{x_{2}} dx \, J_{1,2}(x)  \; .
\label{eq:7a}
\end{equation}
The current $J^{(u)}(x)$ has to satisfy Eq.\ (\ref{eq:5}),
\begin{equation}
\frac{1}{x_{2} - x_{1}}\int_{x_{1}}^{x_{2}} dx \, J^{(u)}(x) = J  \;
.
\label{eq:5x}
\end{equation}
Similarly, Eq.\ (\ref{eq:5e}) is to be replaced with
\begin{equation}
J^{(u)}(x_{1}^{+}) = J^{(u)}(x_{2}^{-}) \equiv \kappa \, J \; .
\label{eq:5y}
\end{equation}
The conditions (\ref{eq:5x}) and (\ref{eq:5y}) determine the coefficients
$\hat{a}_{1}$ and $\hat{a}_{2}$. We find from Eq.\ (\ref{eq:5x}), using
Eqs.\ (\ref{eq:11}) and (\ref{eq:7a}),
\begin{equation}
\hat{a}_{1} + \hat{a}_{2} = 1 \; .
\label{eq:11aa}
\end{equation}
In order to apply condition (\ref{eq:5y}), we evaluate the currents
(\ref{eq:11ax}) and (\ref{eq:11ay}) at the ends of the sample, using Eqs.\
(\ref{eq:6}) and (\ref{eq:7}),
\begin{eqnarray}
\frac{J_{1}(x_{1}^{+})}{J_{1}} &=& w_{1}(x_{1},x_{2};l) \; \chi_{1}(x_{2})
\nonumber \\ &+& \int_{x_{1}}^{x_{2}} \frac{dx'}{l} \, w_{2}(x_{1},x';l) \,
\chi_{1}(x') \; ,
\label{eq:11ax1}
\end{eqnarray}
\begin{eqnarray}
\frac{J_{2}(x_{1}^{+})}{J_{2}} &=& w_{1}(x_{1},x_{2};l) \; \chi_{2}(x_{1})
\nonumber \\ &+& \int_{x_{1}}^{x_{2}} \frac{dx'}{l} \, w_{2}(x_{1},x';l) \,
[\chi_{2}(x_{1}) - \chi_{2}(x')] \; , \nonumber \\
\label{eq:11ax2}
\end{eqnarray}
\begin{eqnarray}
\frac{J_{1}(x_{2}^{-})}{J_{1}} &=& w_{1}(x_{1},x_{2};l) \; \chi_{1}(x_{2})
\nonumber \\ &+& \int_{x_{1}}^{x_{2}} \frac{dx'}{l} \, w_{2}(x',x_{2};l) \,
[\chi_{1}(x_{2}) - \chi_{1}(x')] \; , \nonumber \\
\label{eq:11ax3}
\end{eqnarray}
\begin{eqnarray}
\frac{J_{2}(x_{2}^{-})}{J_{2}} &=& w_{1}(x_{1},x_{2};l) \; \chi_{2}(x_{1})
\nonumber \\ &+& \int_{x_{1}}^{x_{2}} \frac{dx'}{l} \, w_{2}(x',x_{2};l) \,
\chi_{2}(x') \; .
\label{eq:11ax4}
\end{eqnarray}
Employing Eqs.\ (\ref{eq:11}), (\ref{eq:5y}), and
(\ref{eq:11ax1})--(\ref{eq:11ax4}), we obtain the coefficients $\hat{a}_{1,2}$
as given by Eqs.\ (\ref{eq:11bbc})--(\ref{eq:11bfn}).

Now, introducing, in analogy to Eq.\ (\ref{eq:11}),
\begin{equation}
{\cal J}^{(u)}(x) = \hat{a}_{1} \, \frac{J}{J_{1}} \, {\cal J}_{1}(x) +
\hat{a}_{2} \, \frac{J}{J_{2}} \,{\cal J}_{2}(x)
\label{eq:12}
\end{equation}
for $x_{1} < x < x_{2}$, and, in addition,
\begin{equation}
{\cal J}^{(u)}(x_{1}) = \hat{a}_{1} \, \frac{J}{J_{1}} \, {\cal J}_{1} +
\hat{a}_{2} \, \frac{J}{J_{2}} \,{\cal J}_{2}(x_{1}) \; ,
\label{eq:12xx}
\end{equation}
\begin{equation}
{\cal J}^{(u)}(x_{2}) = \hat{a}_{1} \, \frac{J}{J_{1}} \, {\cal J}_{1}(x_2) +
\hat{a}_{2} \, \frac{J}{J_{2}} \,{\cal J}_{2} \; ,
\label{eq:12yy}
\end{equation}
we may write the unique thermoballistic current (\ref{eq:11}), using Eqs.\
(\ref{eq:11ax}), (\ref{eq:11ay}) and (\ref{eq:12})--(\ref{eq:12yy}), in
a symbolic form analogous to expression (\ref{eq:11aba}),
\begin{eqnarray}
J^{(u)}(x) &=&  \int_{x_{1}}^{x} \frac{dx'}{l} \int_{x}^{x_2}
\frac{dx''}{l} \, \mathbb{W}(x',x'';l) \nonumber \\ &\ & \times \left[ {\cal
J}^{(u)} (x') - {\cal J}^{(u)}(x'') \right] \; ;
\label{eq:11abax}
\end{eqnarray}
here, the values ${\cal J}^{(u)}(x_{1,2})$ are to be identified with their
physical values ${\cal J}_{1,2}$ in the contacts,
\begin{equation}
{\cal J}^{(u)}(x_{1,2}) = {\cal J}_{1,2} \; .
\label{eq:16abf}
\end{equation}
In line with the definition (\ref{eq:14a}) of the current ${\cal J}(x)$ in
terms of the chemical potential $\mu(x)$, we now define a unique chemical
potential $\mu^{(u)}(x)$ via relation (\ref{eq:12}) by
\begin{equation}
e^{\beta \mu^{(u)}(x)}  = \frac{1}{v_{e} N_{c}} \, e^{\beta E_{c}^{0}} \,
{\cal J}^{(u)}(x)
\label{eq:12ff}
\end{equation}
for $x_{1} \leq x \leq x_{2}$, where now
\begin{equation}
\mu^{(u)}(x_{1,2}) = \mu_{1,2} \; .
\label{eq:16abe}
\end{equation}
The chemical potential $\mu^{(u)}(x)$ is the key quantity in the extended
unified description of electron transport inside the sample.  In terms of
$\mu^{(u)}(x)$, the unique ballistic current across the interval $[x',x'']$
appearing in the expression for the thermoballistic current (\ref{eq:11abax})
is given by Eq.~(\ref{eq:10xx1}).

For the explicit calculation of the chemical potential $\mu^{(u)}(x)$, we use
Eq.\ (\ref{eq:16abf}) in Eqs.\ (\ref{eq:12xx}) and (\ref{eq:12yy}). We then
find, with the help of Eqs.\ (\ref{eq:6}) and (\ref{eq:7}),
\begin{equation}
\left( 1 - \hat{a}_{1} \frac{J}{J_{1}} \right) {\cal J}_{1} - \hat{a}_{2} \,
\frac{J}{J_{2}} \, {\cal J}_{2} = \hat{a}_{2} \, J  \chi_{2}(x_{1})  \; ,
\label{eq:16abg}
\end{equation}
\begin{eqnarray}
- \hat{a}_{1} \, \frac{J}{J_{1}} \, {\cal J}_{1} + \left( 1 - \hat{a}_{2}
\frac{J}{J_{2}} \right) {\cal J}_{2}  &=& - \hat{a}_{1} \, J
\chi_{1}(x_{2}) \; . \nonumber \\
\label{eq:16abh}
\end{eqnarray}
By subtracting Eq.\ (\ref{eq:16abh}) from Eq.\ (\ref{eq:16abg}), we obtain Eq.\
(\ref{eq:12adac}). On the other hand, adding Eqs.\ (\ref{eq:16abg}) and
(\ref{eq:16abh}) results in
\begin{eqnarray}
\hat{a}_{1} \, \frac{J}{J_{1}} \, {\cal J}_{1}  + \hat{a}_{2} \,
\frac{J}{J_{2}} \, {\cal J}_{2} &=& \frac{1}{2} ({\cal J}_{1} + {\cal J}_{2})
\nonumber \\ &+& \frac{J}{2} [\hat{a}_{1} \, \chi_{1}(x_{2}) - \hat{a}_{2} \,
\chi_{2}(x_{1})] \; .
\nonumber \\
\label{eq:12ijl}
\end{eqnarray}
Then, expressing the current ${\cal J}^{(u)}(x)$ in terms of the quantities
${\cal J}_{1,2}$ and the functions $\chi_{1,2}(x)$ by combining Eq.\
(\ref{eq:12}) with Eqs.\ (\ref{eq:6}) and (\ref{eq:7}),
\begin{eqnarray}
{\cal J}^{(u)}(x) &=& \hat{a}_{1} \, \frac{J}{J_{1}} \, [{\cal J}_{1} - J_{1}
\, \chi_{1}(x)] \nonumber \\ &\ & + \; \, \hat{a}_{2} \, \frac{J}{J_{2}} \,
[{\cal J}_{2} + J_{2} \, \chi_{2}(x)] \nonumber \\ &=&  \hat{a}_{1} \,
\frac{J}{J_{1}} \, {\cal J}_{1} + \hat{a}_{2} \, \frac{J}{J_{2}} \, {\cal
J}_{2} \nonumber \\  &\ & - \; \, J \, [\hat{a}_{1} \, \chi_{1}(x) -
\hat{a}_{2} \, \chi_{2}(x)] \; , \nonumber
\\ \label{eq:12ijk}
\end{eqnarray}
we obtain Eq.\ (\ref{eq:12aci}) with $\chi_{-}(x)$ given by Eq.\
(\ref{eq:12s}). Using Eqs.\ (\ref{eq:12aci}) and (\ref{eq:12adac}) to eliminate
the total current $J$ as well as Eqs.\ (\ref{eq:12ff}) to go over to the unique
chemical potential $\mu^{(u)}(x)$, we find
\begin{eqnarray}
e^{\beta \mu^{(u)}(x)} &=& \left[ \frac{1}{2} - \frac{\chi_{-}(x)}{\chi}
\right] e^{\beta \mu_{1}}  + \left[ \frac{1}{2} + \frac{\chi_{-}(x)}{\chi}
\right] e^{\beta \mu_{2}} \; , \nonumber \\
\label{eq:12ijx}
\end{eqnarray}
where $\chi$ is defined by Eq.\ (\ref{eq:12e}).  The corresponding
thermoballistic current $J^{(u)}(x)$ and density $n^{(u)}(x)$ are
obtained by substituting expression (\ref{eq:12ijx}) in Eq.\ (\ref{eq:11abbc})
and (\ref{eq:11abbd}), respectively.

In the main body of the paper, we always deal with the unique chemical
potential, current, and density, and omit the superscript $u$; we have
already adhered to this convention when referring from the Appendix to the
equations of Sec.\ II.B.

{}

\end{document}